\documentclass[american]{article}
\usepackage[utf8]{inputenc}
\usepackage{authblk}
\usepackage{geometry}
\usepackage{amsmath}
\usepackage{hyperref}
\usepackage{graphicx}
\usepackage{wrapfig}
\graphicspath{ {./Images_nc/} }
\usepackage{subcaption}

\usepackage[dvipsnames]{xcolor}

\usepackage[switch]{lineno}
\usepackage{sectsty}
\usepackage{soul}

\usepackage[sort, round]{natbib}

\begin{document}

\title{Automatic detection of impact craters on Al foils from the Stardust interstellar dust collector using convolutional neural networks}

\author{}

{\centering
\maketitle
Logan Jaeger$^1$, 
Anna L. Butterworth$^1$,
Zack Gainsforth$^1$, 
Robert Lettieri$^1$,
Augusto Ardizzone,
Michael Capraro,
Mark Burchell$^2$,
Penny Wozniakiewicz$^2$,
Ryan C. Ogliore$^3$,
Bradley T. De Gregorio$^4$,
Rhonda M. Stroud$^4$,
Andrew J. Westphal$^1$
\\\ \\
\centerline{$^1$Space Sciences Laboratory, University of California at Berkeley, Berkeley, CA 94720 USA}
\centerline{$^2$Centre for Astrophysics and Planetary Science, University of Kent, Canterbury CT2 7NH, UK}
\centerline{$^3$Washington University, St. Louis, MO, 63130, USA}
\centerline{$^4$Materials Science and Technology Division, Code 6366, US Naval Research Laboratory, Washington, DC 20375–5320, USA}
}

\newpage

\begin{abstract}
    NASA's Stardust mission utilized a sample collector composed of aerogel and aluminum foil to return cometary and interstellar particles to Earth. 
    Analysis of the aluminum foil begins with locating craters produced by hypervelocity impacts of cometary and interstellar dust.  
    Interstellar dust craters are typically less than one micrometer in size and are sparsely distributed, making them difficult to find.  
    In this paper, we describe a convolutional neural network based on the VGG16 architecture that achieves high specificity and sensitivity in locating impact craters in the Stardust interstellar collector foils. 
    We evaluate its implications for current and future analyses of Stardust samples.
\end{abstract}

\sectionfont{\centering}

\section*{INTRODUCTION} 
\begin{figure}[p]
    \centering
    \begin{subfigure}[t]{.3 \textwidth}
        \includegraphics[width=.95\linewidth]{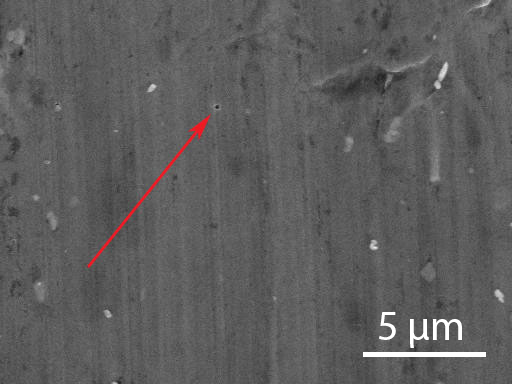}
        \caption{}
        \label{fig:crater_1}
    \end{subfigure}
    \begin{subfigure}[t]{.3 \textwidth}
        \includegraphics[width=.95\linewidth]{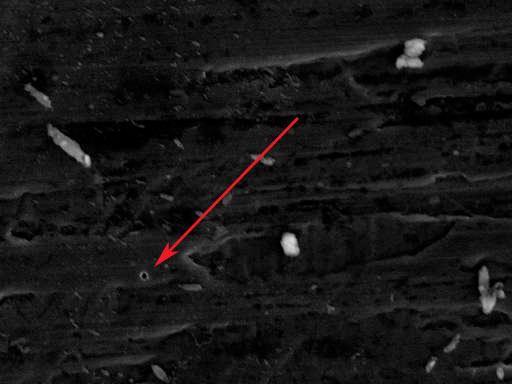}
        \caption{}
        \label{fig:crater_2}
    \end{subfigure}
        \begin{subfigure}[t]{.3 \textwidth}
        \includegraphics[width=.95\linewidth]{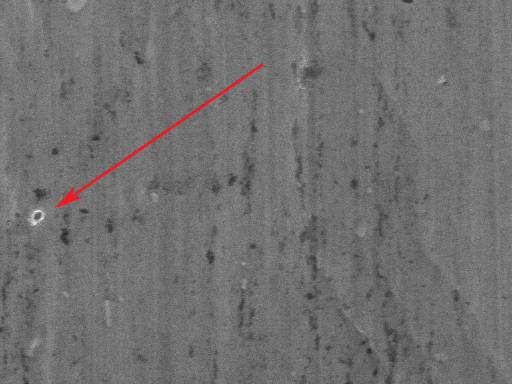}
        \caption{}
        \label{fig:crater_3}
    \end{subfigure}
    
    \caption{Some examples of images that contain real craters from the Stardust interstellar collector. Notice the differences in size, shape, and image exposure. (a) and (c) come from foil I1020W, while (b) comes from foil I1126N.}
    \label{fig:craters}
\end{figure}

\begin{figure}[p]
    \centering
    \begin{subfigure}[t]{.3\textwidth}
        \includegraphics[width = .95\linewidth]{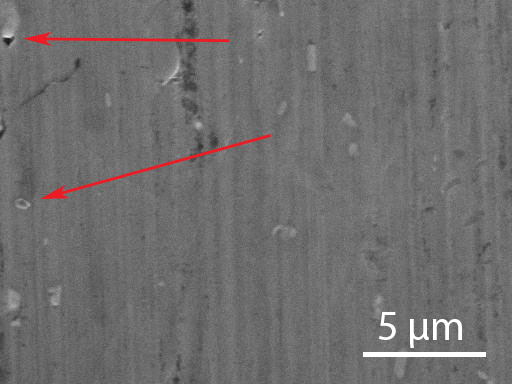}
        \caption{}
        \label{fig:NoCrater1}
    \end{subfigure}
    \begin{subfigure}[t]{.3\textwidth}
        \includegraphics[width = .95\linewidth]{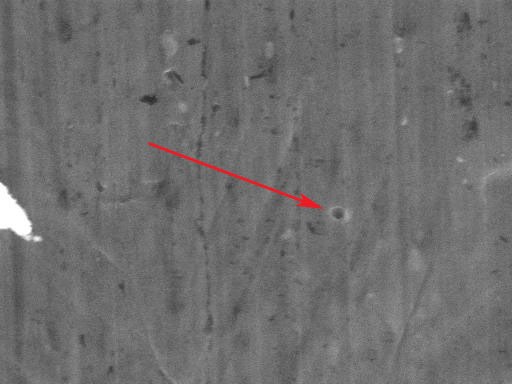}
        \caption{}
        \label{fig:NoCrater2}
    \end{subfigure}
    \begin{subfigure}[t]{.3\textwidth}
        \includegraphics[width = .95\linewidth]{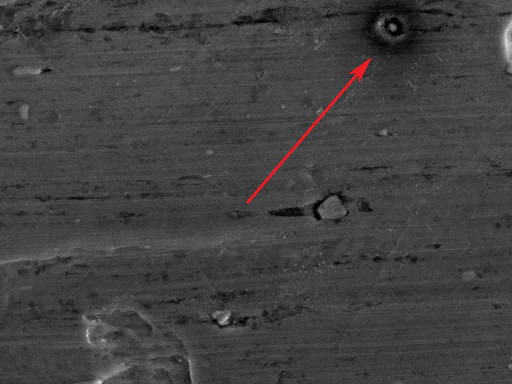}
        \caption{}
        \label{fig:Sulfide}
    \end{subfigure}
    
    \caption{Examples of images that do not contain craters, but do contain some misleading imperfections in the foil. These features are either impurities introduced during the manufacturing process or are impacts produced by secondary ejecta from the Stardust spacecraft \citep{burchell2012experimental}. (c) contains a iron oxide ``volcano", which occurs when there is a iron impurity near the surface of the foil that ``erupts" out. (a) and (b) are taken from foil I1020W and (c) is from foil I1126N.}
    \label{fig:not_craters}
\end{figure}

NASA's Stardust mission successfully returned cometary and interstellar dust collectors to Earth in 2006, after an encounter with the comet 81P/Wild 2 in 2004, and exposure to the interstellar dust stream in 2000 and 2002. 
The cometary and interstellar dust particles it collected have been extensively studied, leading to scientific insights on the origins of the solar system \citep{brownlee2006comet, brownlee2014review}, the nature of comets \citep{brownlee2004surface, mckeegan2006isotopic, flynn2006elemental, zolensky2006mineralogy, simon2008refractory, joswiak2012comprehensive, ogliore2012incorporation, gainsforth2019fine}, the presence of cometary amino acids \citep{elsila2009cometary, sandford2006organics}, and the properties of interstellar dust \citep{westphal2014evidence, stroud2014stardust, bechtel2014stardust, brenker2014stardust, flynn2014stardust, butterworth2014stardust, gainsforth2014stardust, westphal2014final}. 
Stardust carried one collector to capture cometary particles from 81P/Wild 2 and another to capture  interstellar dust particles.
Each collector consisted of aerogel tiles and aluminum foils. The foils were intended to facilitate aerogel removal, but also provided an additional capture medium \citep{tsou2003wild}.
When particles hit the aluminum foils, they created small craters \citep{horz2006impacts, kearsley2008dust, price2010, wozniakiewicz2012stardust}.
These craters are sparsely distributed on the foils and the majority are only detectable by scanning electron microscope (SEM) imaging.
Even then they can easily be confused with imperfections in the foil (See Fig. \ref{fig:craters} and Fig. \ref{fig:not_craters}), making the task of locating them nontrivial.

A large citizen science project, Stardust@home (hereafter SAH), engaged human volunteers to locate craters on the Stardust foils \citep{westphal2005stardust}. 
This project has attracted more than 30,000 volunteers and has been particularly successful with respect to the identification of particles in aerogel \citep{westphal2014stardust}. 
However, even with this massive participation, human views of the foils are still sparse.
Nearly 20\% of the available SAH foil images have not been viewed more than once.
Furthermore, the SAH site only contains images from four of the roughly 200 individual foils present in the Stardust collector.
More importantly, the majority of the foils in the Stardust collector have never been scanned by a SEM. \cite{stroud2014stardust} reports that approximately 4.8\% of the foils were examined by a SEM during the preliminary examination. 
Since then, just over half of the foils have been removed from the collector. 
An efficient and effective automated procedure could provide an impetus to scan the rest of these foils, as the examination of images would be much easier.

Previous successful attempts at automatic crater identification include a normalized cross-correlation and template matching algorithm by \cite{ogliore2012automated} and an algorithm based on a circular Hough transform and Canny edge detection developed by \cite{de2018fast}. 
Both of these algorithms have revealed multiple new craters. 
Convolutional neural networks (CNNs), which have been established as effective tools in image classification \citep{he_delving_2015, krizhevsky_imagenet_2012}, have the potential to complement or outperform the previously developed machine learning techniques. 

\begin{figure}[p]
	\centering
	\begin{subfigure}[t]{.45\textwidth}
		\includegraphics[width = .95\linewidth]{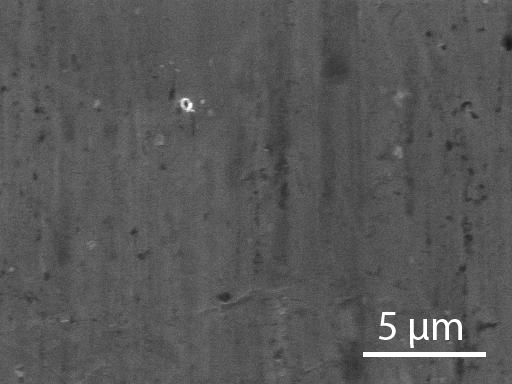}
		\caption{}
	\end{subfigure}
	\begin{subfigure}[t]{.45 \textwidth}
		\includegraphics[width = .95\linewidth]{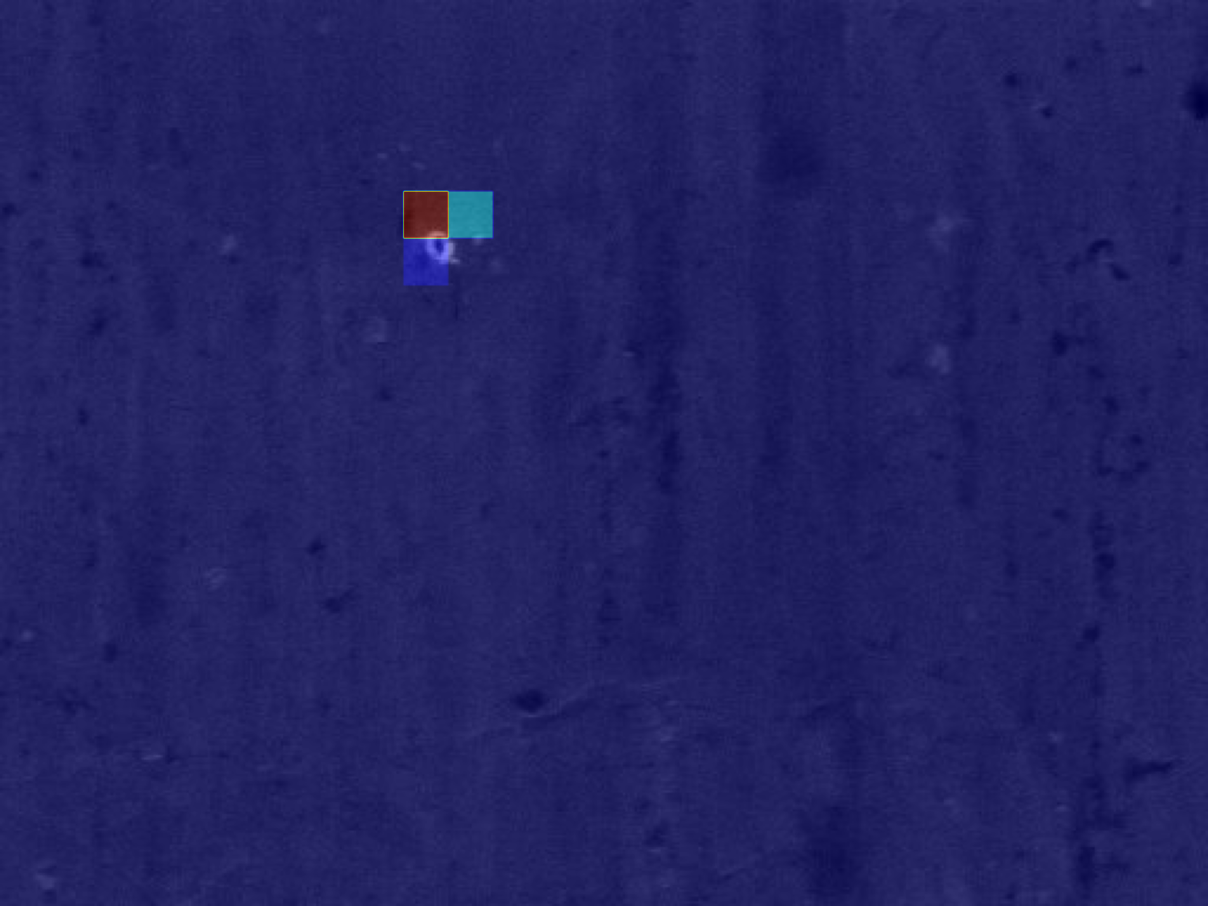}
		\caption{}
	\end{subfigure}
		\begin{subfigure}[t]{.45\textwidth}
		\includegraphics[width = .95\linewidth]{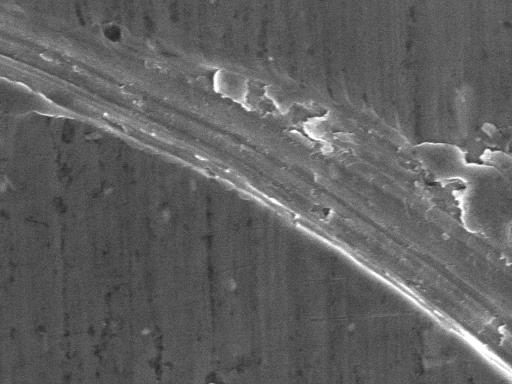}
		\caption{}
	\end{subfigure}
	\begin{subfigure}[t]{.45 \textwidth}
		\includegraphics[width = .95\linewidth]{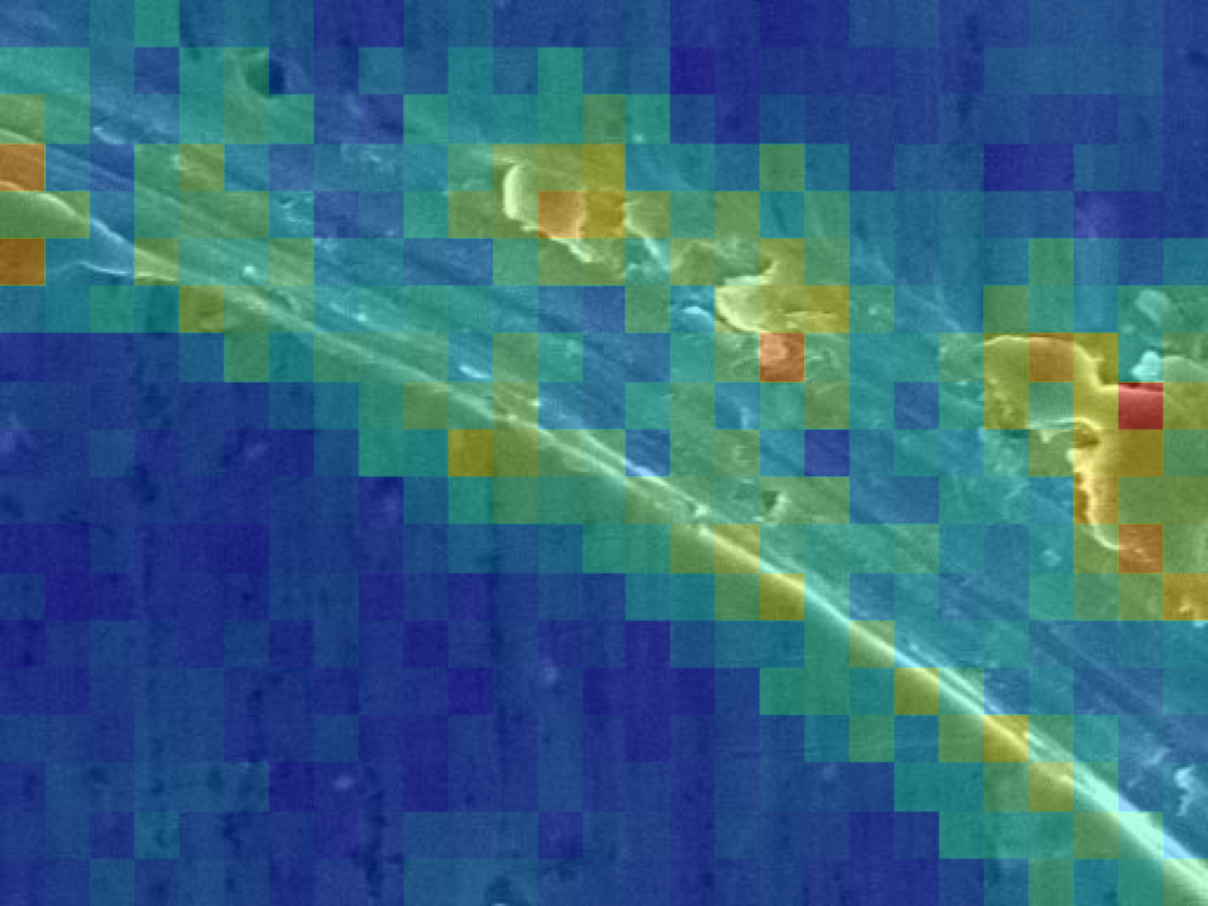}
		\caption{}
	\end{subfigure}
    \caption{Heatmaps created by overlaying a scaled-up activation map from one of the filters on the final convolutional layer over the input image. These reveal an advantage of a CNN over previous attempts at crater identification. Earlier methods only ``trigger" on features that look like craters without accounting for the surrounding conditions. A CNN, on the other hand, contains many filters in each convolutional layer, with each filter having a different application. Some filters are focused on identifying craters, like that in (b), while others can see features like the scratch identified by the filter in (d). In this way, a CNN can account for the relationship between a crater-like object and its environment to better determine whether or not it truly is a crater. Both images are from the foil I1020W.}
	\label{fig:CNNHeatmaps}
\end{figure}

By way of comparison, the Hough transform algorithm detects circular craters, and template matching detects craters matching a set of known craters.
A CNN, on the other hand, can be trained on craters in combination with other features, such as scratches and image exposure, and can integrate those observations into the final metric.
Thus, while both the Hough transform algorithm and the cross-correlation template matching algorithm only seek to find crater-like features, a CNN can both identify crater-like features and address the environment of a given image. (See Fig. \ref{fig:CNNHeatmaps}).
Because the fraction of SEM images that contain craters is very small, Bayes' theorem implies that the majority of images classified as containing craters by our algorithm will actually be false positives, even with a generous estimation of the neural net accuracy. \
For example, let us we suppose that our CNN achieves a value of 0.999 for both sensitivity and specificity, and we take the estimate that there there are 100 total craters in the sample of approximately $10^7$ images required to search the entire collector \citep{landgraf1999prediction}. Even so, Bayes' theorem says there is still less than a one percent chance that an image classified to contain a crater actually contains a crater. 
Where $P(\textrm{c})$ represents the probability that a crater is actually present in an image, $P(\overline{\textrm{c}})$ represents the probability that a crater is \textit{not} present in an image, and $P(+)$ represents the probability that our CNN \textit{predicts} a crater is present in an image:

\begin{equation*}
\begin{split}
    P(\textrm{c}|+) &= \frac{P(+|\textrm{c})P(\textrm{c})}{P(+|\overline{\textrm{c}})P(\overline{\textrm{c}}) + P(+|\textrm{c})P(\textrm{c})} \\ 
    &= \frac{.999 \times (100/10^7)}{.001 \times (1-(100/10^7)) +  .999 \times (100/10^7)} \\ 
    & \approx 1\% \textrm{ probability a crater is present in a positively identified image}
\end{split}
\end{equation*}

So a significant amount of human interaction will still be necessary to ultimately find and identify craters, even if we reach the bounds of practical CNN performance. 
The focus of the project, then, is not to make a perfect crater identification algorithm that eliminates the need for human interaction with these images.  
Instead, we intend to make an algorithm that is good enough to correctly identify the large majority of images that \textit{do not} contain craters, thereby reducing the amount of time and effort humans must expend looking through images. 
That is, we are biasing our CNN towards being highly specific, rather than being highly sensitive.

Cometary particles are generally larger than interstellar particles, and create craters with a typical diameter of $5-200$ $\mu$m \citep{kearsley2008dust}, as opposed to a diameter of $< 1$ $\mu$m for typical interstellar craters \citep{stroud2014stardust}.
However, there is interest in smaller cometary particles as well, as $\mu$m to sub-$\mu$m cometary craters have been examined by \cite{leroux2008transmission}, \cite{stroud2010structure} and \cite{Leitner}.
All of the data used in this paper comes from the interstellar dust collector, but the results could be applied to the cometary collector.


\section*{EXPERIMENTAL} 

\subsection*{SAMPLE PREPARATION} 
The top foil surface was separated from the collector apparatus using a twin rotary cutter in the Stardust Lab at Johnson Space Center.  This resulted in thin Al strips that were stored and shipped in a custom foil mount \citep{kearsley2008dust}. 
In preparation for SEM imaging, we clamped the foils mechanically at the ends and stretched them to ensure flatness.
This method of foil mounting was also described in \cite{stroud2014stardust} and has been shown to ensure stability and meet the standards necessary for automatic image acquisition. 
We imaged the foils in SEMs using voltage between 5 keV and 15 keV, current between 0.2 nA and 5 nA, and image resolution between 40 and 60 nm px$^{-1}$ \citep{stroud2014stardust}.
Care was taken to ensure that imaging times were sufficiently short, and that the vacuum chambers within the SEMs used were sufficiently free of carbon, in order to minimize deposition that could compromise later investigation.
For example, Auger spectroscopy can be used for elemental mapping of craters in the Stardust foils but is extremely sensitive to carbon deposition and can be obscured even by a 4 nm layer of carbon \citep{stadermann2009use}.

Due to the small number of actual craters that have thus far been observed in the SEM images of the Stardust collector, this project would be impossible without analog craters produced by alternative means, namely light gas gun shots of aluminum foils \citep{burchell1999hypervelocity}. 
44 such craters had originally been imaged for use by \cite{stadermann2008stardust}, and had previously been used by \cite{ogliore2012automated}.
We produced a further 53 with the two-stage light gas gun (LGG) in the Centre for Astrophysics and Planetary Sciences at the University of Kent (see \citealt{burchell1999hypervelocity}), using the procedure described in \cite{wozniakiewicz2012stardust}. 
The method involved firing projectiles at an analog Stardust aluminium foil using a two stage light gas gun. 
The projectiles were a mixed sample, comprising 0.5 and 1.0 $\mu$m diameter SiO$_2$ spheres, mixed with larger grains of hand-ground pyroxene, olivine and diopside. 
We measured the impact speed to be 6.02 $\pm$ 0.05 km s$^{-1}$, comparable to the Stardust cometary dust impact speed. 
The craters from the smaller SiO$_2$ grains were identifiable by their regular circular shape and size. 
Based on \cite{price2010} it is predicted that, for silica beads similar to those here impacting Al foil at 6 km s$^{-1}$, crater size = 1.54 times projectile diameter. 
Hence crater sizes of 0.77 $\mu$m and 1.54 $\mu$m are expected from the SiO$_2$ spheres, as well as larger craters from the mineral impacts. The larger craters from the various minerals would have distinct differences in residue composition inside the craters. 
We imaged the foil using a Tescan Vega SEM at the Space Sciences Laboratory in Berkeley. 


\subsection*{DATA PREPROCESSING}
\begin{figure}[p]
    \centering
        \begin{subfigure}[t]{.45 \textwidth}
        \includegraphics[width=.95\linewidth]{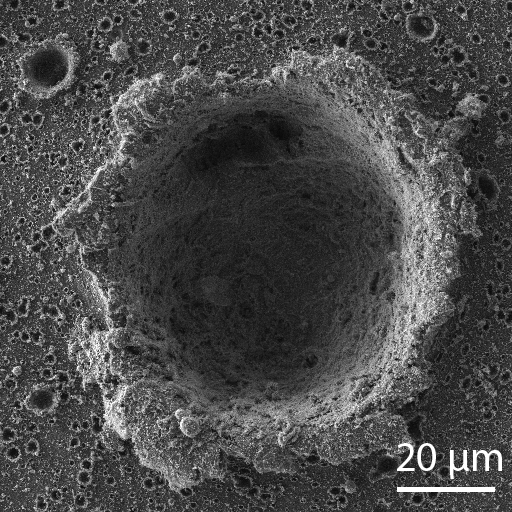}
        \caption{}
        \label{fig:crater-with-back}
    \end{subfigure}
    \begin{subfigure}[t]{.45 \textwidth}
        \includegraphics[width=.95\linewidth]{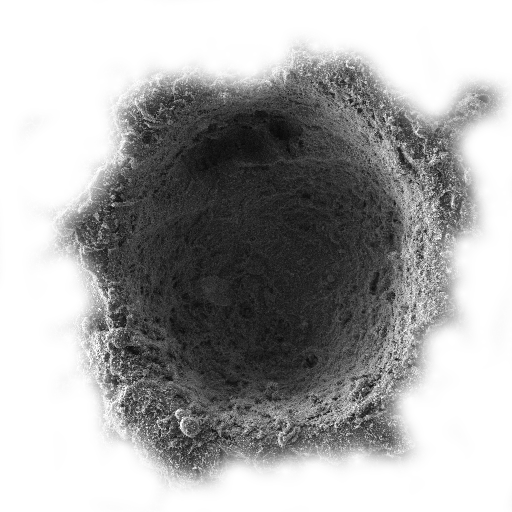}
        \caption{}
        \label{fig:alpha-crater}
    \end{subfigure}
    \begin{subfigure}[t]{.45 \textwidth}
        \includegraphics[width=.95\linewidth]{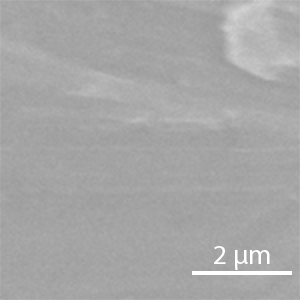}
        \caption{}
        \label{fig:blank-150150}
    \end{subfigure}
    \begin{subfigure}[t]{.45 \textwidth}
        \includegraphics[width=.95\linewidth]{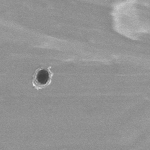}
        \caption{}
        \label{fig:pasted-150150}
    \end{subfigure}

    \caption{The process of augmenting images. A crater is imaged by hand with a SEM (a), then isolated from the background using a soft brush (b). A SEM image of a Stardust foil that does not contain a crater (c) is the background into which the isolated crater is pasted (d), after having been augmented. Here, the SEM image of the stardust foil is $150 \times 150$ pixels, of the type used in the first step of CNN training. The stigmation in the crater image is slightly different than the stigmation in the background image, which will create a slight bias when we train our network, as a true image that contains a crater would have the exact same stigmation for both the crater and the background. In future versions of the network, we plan to adjust the stigmation of the crater to match that of the background.}  
    \label{fig:image-augmenting}
\end{figure}

Adequate CNN training requires a set of thousands of images that contain craters and thousands that do not. 
While we have many images without craters, images with craters in the interstellar collector are rare.
We resolved this issue by augmenting a set of blank images with analog craters that sample the range of crater morphologies, as described in the section above. 
The process is shown in Fig. \ref{fig:image-augmenting}.  A crater is first imaged (Fig. \ref{fig:crater-with-back}), isolated from the surrounding foil (Fig. \ref{fig:alpha-crater}) and then combined with a blank image (Figs. \ref{fig:blank-150150} and \ref{fig:pasted-150150}).
In order to isolate the crater from the surrounding foil, we hand-edited each image to delete the surrounding foil using a transparency channel.
We did this using the GIMP digital image processing package \citep{gimp} with brushes of varying sizes and hardnesses between 25 and 100 pixels in order to reduce the effect of a hard boundary at the edge of the crater or sensitivity to fourier components harmonic with the brush size.  We then performed a process of augmentation of these craters through rotating, re-scaling, and altering the aspect ratio to produce about 30,000 artificial crater images \citep{scikit-image}.
The aspect ratio is sampled from a normal centered at 1 with a standard deviation of 0.1, and the re-scaling is such that a crater is no smaller than 5 pixels in diameter and no larger than about 25, matching the observed scale.
Such data augmentation techniques, known collectively as data warping, are well established in the field of deep learning and computer vision \citep{shorten2019survey, yu2017deep, wong2016understanding}. 
We then pasted these augmented craters into foil images that do not contain craters. We visually checked that these synthetic images looked reasonable. (Fig. \ref{fig:pasted-150150}). 
Before training, we normalized the images such that their pixel intensities were between 0 and 1.
Finally, we split the synthetic images into training, validation, and test sets of 10,000 each, with a similar set of blank images for a total database of 60,000 images.


\subsection*{NETWORK OVERVIEW} 

\begin{figure}[p]
	\centering
	\includegraphics[width=.9\linewidth]{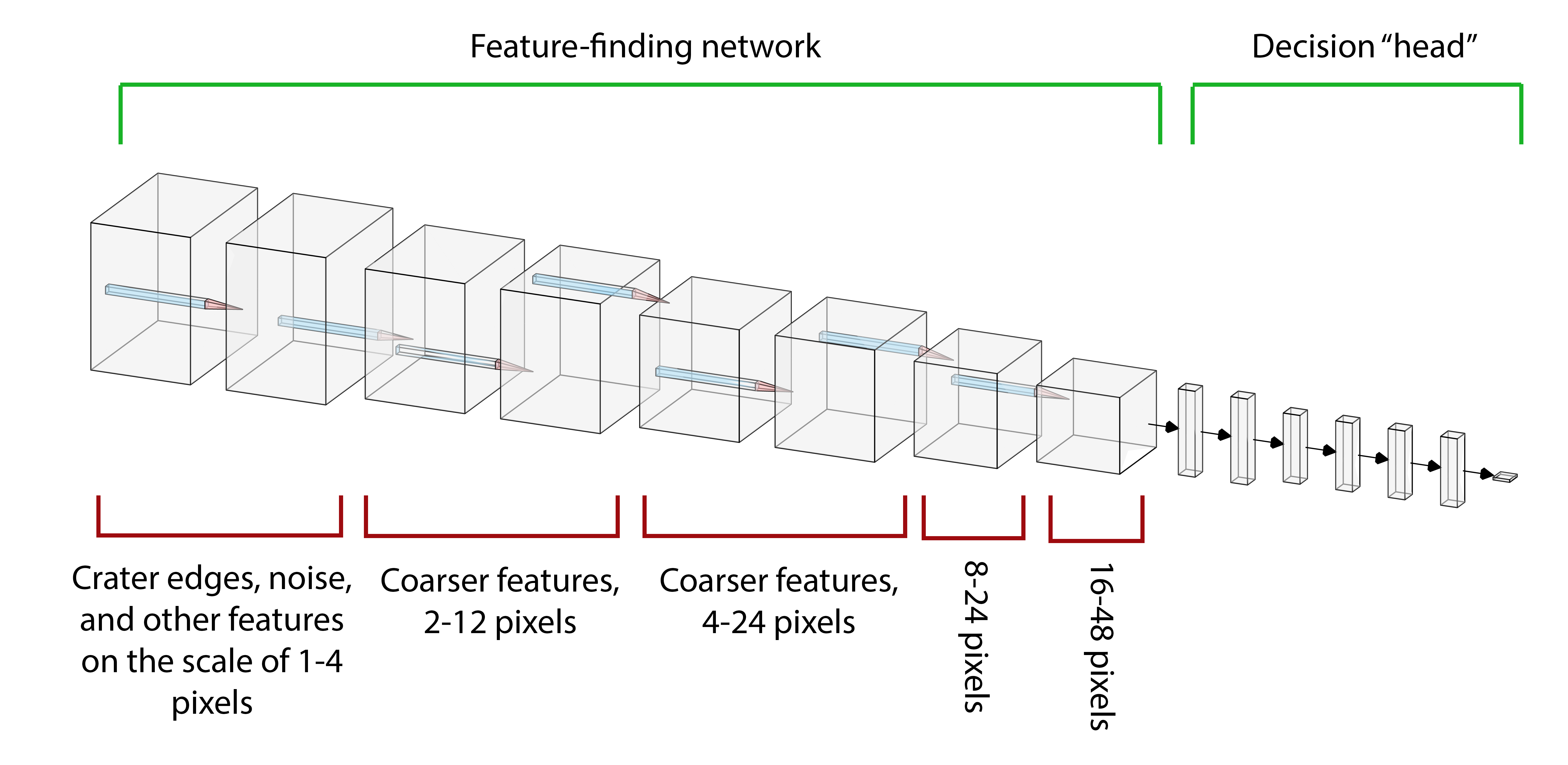}
	\caption{Overview of the convolutional neural network structure showing the 5 feature finding units (8 layers) and the 7 layer decision portion. (Visualization using \citealt{lenail2019nn}.)}
	\label{fig:CNNStructure}
\end{figure}

Our network structure is similar to the VGG16 network \citep{simonyan2014very} frequently used in the machine vision community, but is tailored to our specific application.  
The most significant difference is that our network is substantially smaller than VGG16, having $<$250,000 free parameters, while VGG16 has $>$10$^8$ free parameters.
Figure \ref{fig:CNNStructure} shows an overview of our network structure.  
We used a set of three coarse feature finding units, each with two convolutional layers and a max pooling layer \citep{giusti2013fast}.
These are followed by two more convolutional layers each with a max pooling layer, before feeding into a fully connected head.  
The coarse feature finding units identify features up to about 24 pixels wide, which is the size of a typical crater in our images.  
The additional two convolutional units provide context up to 48 pixels across.  
During hyperparameter optimization we found that features larger than $\sim$50 pixels across contributed little to determining the presence of a feature that looks like a crater.
Thus, we do not include convolutional layers for larger feature sizes.  
The head combines the features found by the convolutional filters to determine whether a crater is present or not.  


\subsection*{DETAILED ARCHITECTURE} 
We programmed the CNN in Keras \citep{chollet2015keras} with a Tensorflow backend \citep{tensorflow2015-whitepaper} using python. 
We utilized a number of other packages in image preprocessing and evaluation, including matplotlib \citep{Hunter:2007}, scikit-learn \citep{scikit-learn}, scikit-image \citep{scikit-image}, hdf5 \citep{hdf5}, numpy \citep{5725236}, and Jupyter notebooks \citep{kluyver2016jupyter}. 

A tabular representation of the network architecture can be seen in Table \ref{table:network}. For brevity, we have only included the main layers in Table \ref{table:network}.  
The CNN had a total of 216,465 trainable parameters, 8 two-dimensional convolutional layers and 7 densely connected layers. 
All hidden layers are immediately followed by the ReLU activation function \citep{krizhevsky_imagenet_2012}, while the output layer is equipped with a sigmoid activation function. 
Every convolutional layer uses a $3 \times 3$ kernel size, with a stride of 1 (stride is the number of pixels between the centers of adjacent kernel operations), and does not use padding. 
All max pooling layers use a $2 \times 2$ pool-size and a stride of 2 pixels. 
We used several methods to prevent overfitting: Following every convolutional layer, there is a two-dimensional spatial dropout layer \citep{Tompson2015EfficientOL} with a dropout rate of 0.25. 
Following every densely-connected layer, there is a simple dropout layer \citep{JMLR:v15:srivastava14a} with a dropout rate of 0.3. 
Every convolutional layer is equipped with a ridge-regression L2-loss kernel regularizer with regularization parameter $\lambda = 0.0001$ \citep{krogh1992simple}.  
As a bridge between the feature-finding and densely-connected decision-making sections of the network, we used a global max pooling layer, as opposed to a flatten layer.
A global max pooling layer not only allows a variable-size input, but also helps prevent overfitting by minimizing the amount of noise passed to the decision-making section \citep{lin2013network}.


\subsection*{NETWORK TRAINING} 

The raw data consist of $384 \times 512$ pixel grayscale JPG images. 
It is computationally costly to train on images of that size. 
To alleviate this cost, we allowed our network to accept a variable-size grayscale input, and trained in multiple steps. 
In the first step, we trained on up to 20,000 synthetic $150 \times 150$ pixel grayscale images (10,000 each for positive and negative images). 
In the second step, we retrained the network, starting from the previously obtained weights, on up to 20,000 grayscale images sized $384 \times 512$ pixels.
In both steps, we used the testing and validation sets of up to 20,000 images each. 

For our optimizer we used Adam with Nesterov momentum \citep{Dozat2016IncorporatingNM}, with a constant learning rate of 0.002. 
We used binary cross-entropy for our loss function.  
In order to bias the network against false positives, we adjusted the class weights such that a false positive was penalized 10 times more than a false negative. 


\subsection*{BIASING THE TRAINING SETS}  

Performance can often be increased by ensemble learning (e.g. \cite{huang2017snapshot}, \cite{caruana2004ensemble}, \cite{dietterich2000ensemble}).
Our current prediction algorithm is in fact an ensemble of three networks each trained with slightly different training data.
They mostly identify the same images.
However, some edge cases (ambiguous features that \textit{may} be craters) are found by only one or two of the networks in the ensemble and therefore we capture more interesting craters by using the union of the predictions.

We produced different versions by varying the training sets as shown in Table \ref{table:Versions}. 
There are two main differences that separate the versions: the number of images used in training and the content of training images. 
The impact of using more images is clear: larger training sets resulted in more accurate networks after training.
However, while V3 and V4 are nominally trained using the same number of images, V4 was trained using only synthetic images produced with our analog craters using the process described above, whereas V2 and V3 were trained with images from the SAH site that had been biased by human volunteer responses.

The negative images for the V2/V3 training sets were chosen from images that had been consistently marked negatively by the SAH users (fewer than than 33\% of the views had indicated a crater).  
Positive images were chosen from SAH ``calibration" images, which are used to measure volunteer performance and had been previously manufactured at the launch of the SAH site in a similar manner described by \cite{westphal2014stardust}, that had been consistently marked \textit{positively} by SAH users (greater than 50\% had indicated a crater). 
Note that all of these calibration images contain synthetic craters, and the volunteer bias was used so that we could be sure that the craters were clear and realistic.
This approach was used to ensure that our network could mirror the activity of SAH volunteers, who have been established as largely reliable in the aggregate \citep{westphal2014stardust}.
These ``calibration" craters were manufactured using only a few analog craters which were all largely circular, and the use of more analog craters with a larger variety of structures helped V4 achieve higher sensitivity and specificity than V2 or V3.


\section*{RESULTS} 

An overview of characteristics and performance of each CNN version is shown in Table \ref{table:Versions}. 
The specificity was measured on a set of 500 images that do not contain craters, chosen randomly from the SAH database. 
Two different measures of sensitivity are presented so as to evaluate both the CNN and our method of synthetic crater generation. The sensitivity with synthetic craters was measured on a set of 500 images that had synthetic craters, reflective of the way in which the CNNs were trained.
That is, the dataset used for V2 and V3 contained the SAH calibration images, and the dataset used for V4 contained synthetic craters created in the manner described in the data preprocessing section. 

When an image from the SAH database is examined by our team and is determined to be a crater candidate, that corresponding section of the foil is re-scanned to obtain a higher resolution image. 
Not all of these crater candidates that are re-scanned are determined to be of interstellar origin (some are crater-looking foil defects, others are craters created by spacecraft ejecta).
Our CNN should nevertheless identify these features as craters because, without a higher resolution image or elemental analysis, distinguishing them from interstellar craters is not possible.
The sensitivity with true craters is given as the proportion of 27 images verified by any and all means over the years including higher resolution SEM imageing, and FIB/TEM.  
Each was verified by further study to be a real impact crater.  
It is apparent that the V4 version of the network, which incorporates the most advanced image augmentation techniques, performed the best at finding features that were later verified to be valid craters.

The normalized cross-correlation algorithm of \citet{ogliore2012automated} (\url{https://presolar.physics.wustl.edu/Laboratory_for_Space_Sciences/software/cratersource.zip}) identified 8 of 25 true craters, indicating that its sensitivity is higher than V3 but lower than V4 of the network. However, the specificity is significantly lower, with 1--3 false positives for most images. Each image took a comparatively long time to search: about four minutes per processor core. Clearly, the network provides a significant improvement in sensitivity, specificity, and speed over a conventional normalized cross-correlation approach.

\begin{figure}[p]
	\centering
	\begin{subfigure}[t]{1 \textwidth}
		\includegraphics[width = .95\linewidth]{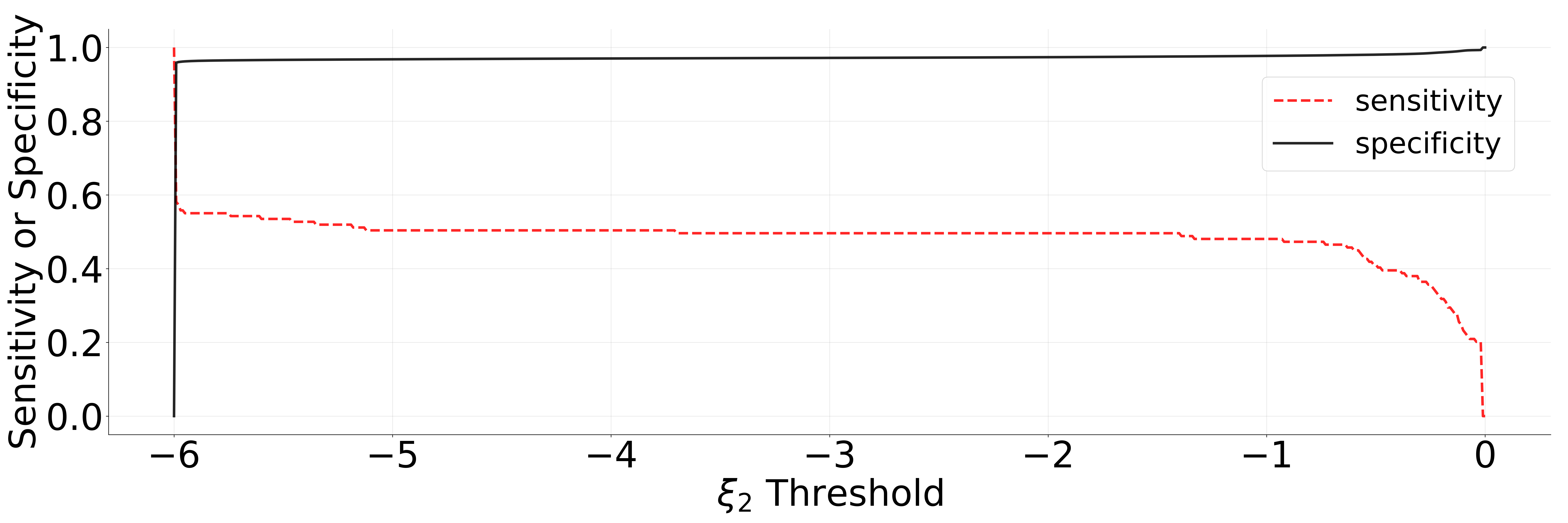}
		\caption{}
		\label{fig:V2 sense spec}
	\end{subfigure}
	\begin{subfigure}[t]{1 \textwidth}
		\includegraphics[width = .95\linewidth]{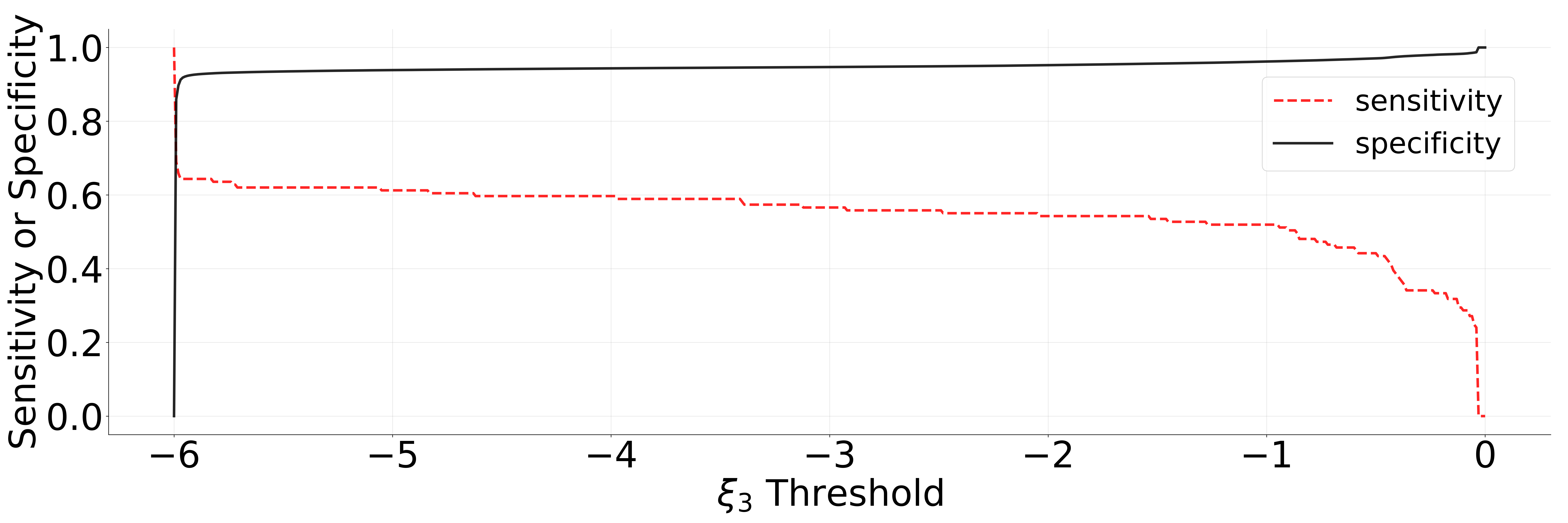}
		\caption{}
		\label{fig:V3 sense spec}
	\end{subfigure}
	\begin{subfigure}[t]{1 \textwidth}
		\includegraphics[width = .95\linewidth]{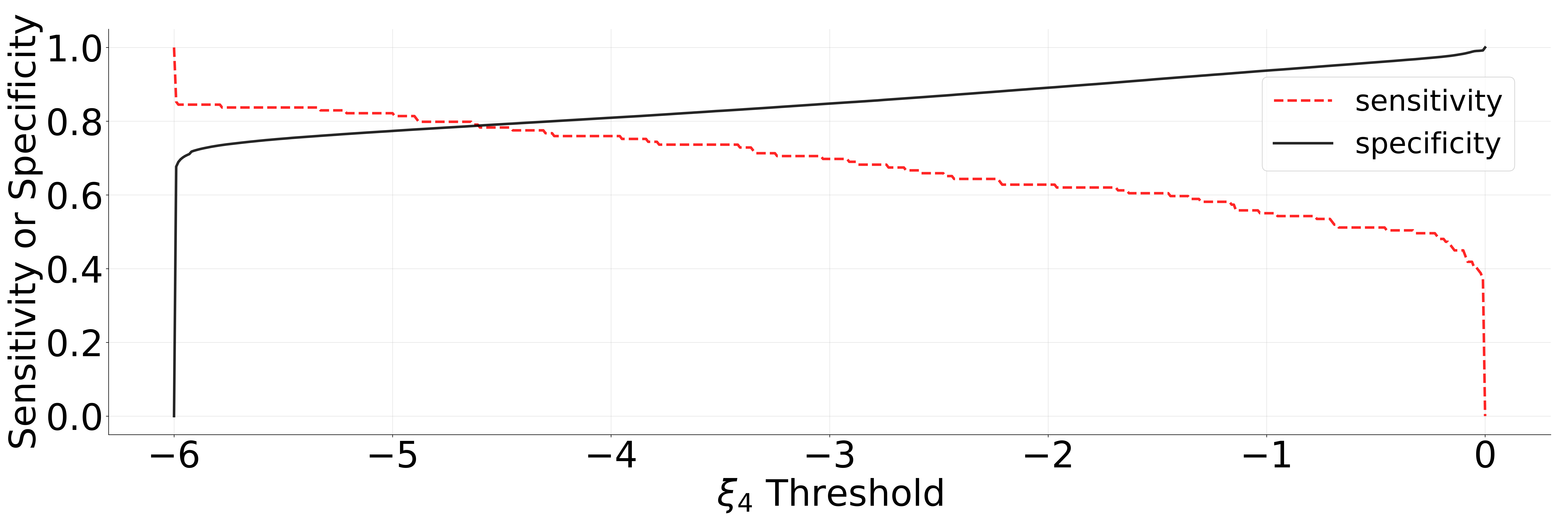}
		\caption{}
		\label{fig:V4 sense spec}
	\end{subfigure}
	\caption{Plots of sensitivity and specificity as a function of threshold. From top to bottom, the graphs correspond to V2, V3, and V4. Sensitivity is calculated on 129 images identified as especially likely to contain a crater by experienced SAH volunteers.}
	\label{fig:SensAndSpec}
\end{figure}

The output of the CNN is not binary, but is a continuous variable, known as the ``prediction", which is related to the probability that a crater is present.
In Fig. \ref{fig:SensAndSpec}, we plotted the sensitivity and specificity of each version of the network as a function of the minimum CNN prediction.
Because much of the change in specificity and sensitivity occurs near a prediction of zero, we define a new parameter $\xi$ as 

$$\xi \equiv \log_{10} (p+10^{-6})$$

where $p$ is the prediction of a given version of a the network.
A small constant, $10^{-6}$, has been added to the prediction inside the logarithm so that a prediction of zero does not diverge.
This new parameter allows us to ``zoom in" on prediction values near zero.
The sensitivity in Fig. \ref{fig:SensAndSpec} is calculated on 129 images that have been selected as especially likely to contain craters by two of us, AA and MC.
Notice that the specificity is high for any threshold greater than zero, and climbs relatively slowly after that, while the sensitivity continues to drop steadily. 
The equations used to calculate the sensitivity and specificity are shown below, where TP, TN, FP, and FN represent the numbers of true positives, true negatives, false positives, and false negatives, respectively.

\begin{gather*}
	\textrm{sensitivity} = \frac{\textrm{TP}}{\textrm{TP} + \textrm{FN}} \qquad
	\textrm{specificity} = \frac{\textrm{TN}}{\textrm{TN} + \textrm{FP}}
\end{gather*}

\begin{figure}[p]
	\centering
	\includegraphics[width = \linewidth]{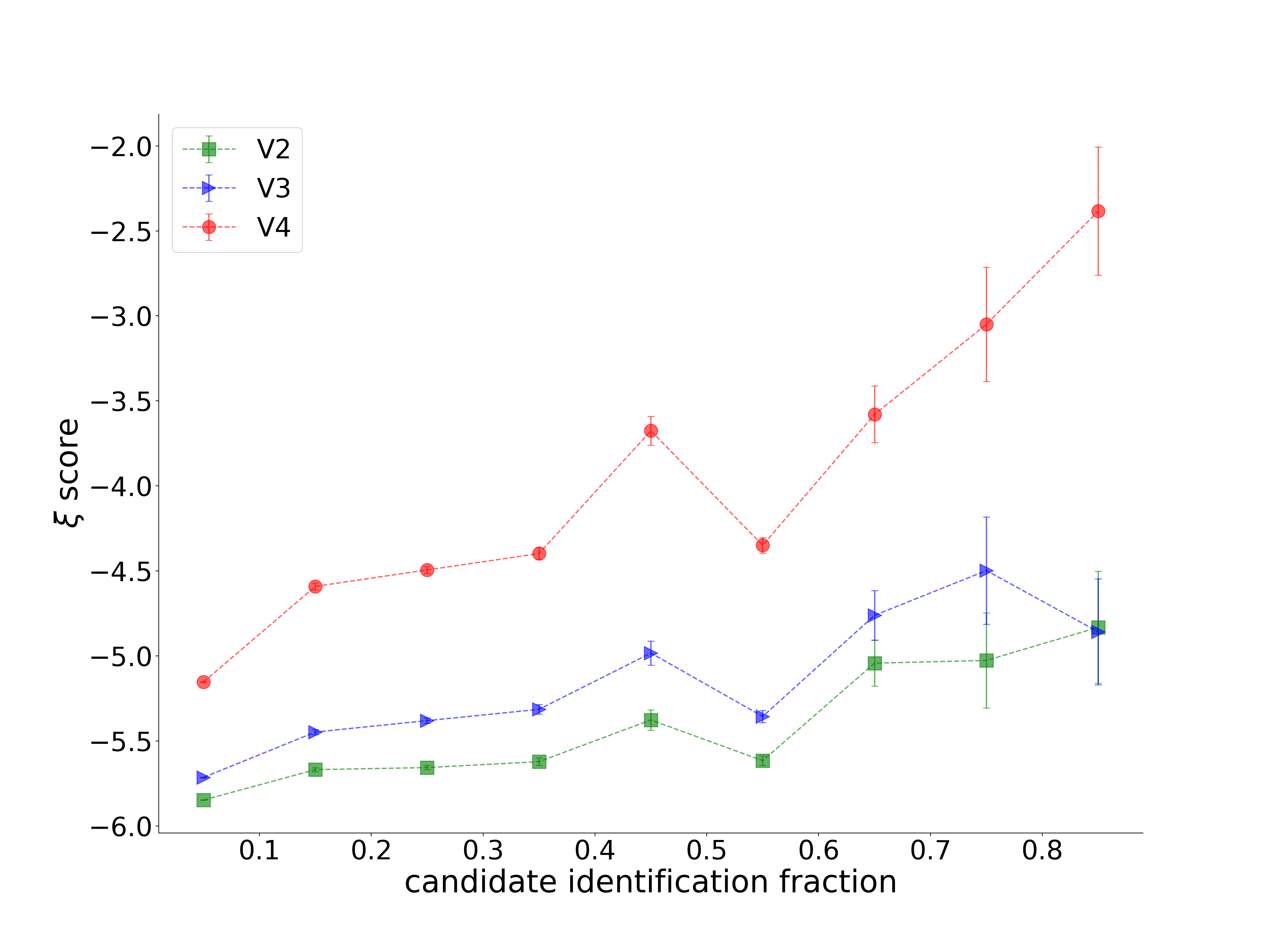}
	\caption{The average $\xi$ score versus crater identification fraction. The crater identification fraction for a given image is the number of times a SAH volunteer has indicated a crater is likely present in the image, divided by the total number of times that the image has been seen by SAH volunteers. }
	\label{fig:graph-probs}
\end{figure}

Because SAH volunteers have been established as largely reliable, we can use their responses as a proxy measurement for the success of the network. 
Fig. \ref{fig:graph-probs} shows that there is a clear correlation between SAH volunteer responses and the predictions of the network for all three versions, with V4 being the most successful.

\begin{figure}[p]
	\centering
	\begin{subfigure}[t]{1 \textwidth}
		\includegraphics[width = .95\linewidth]{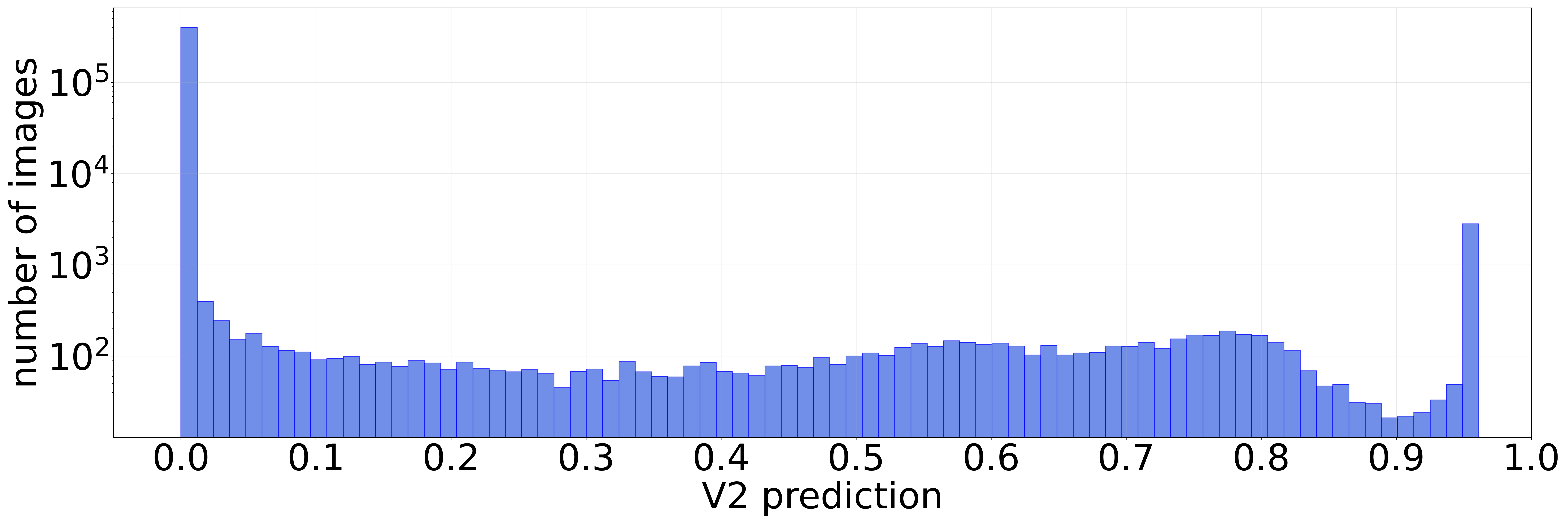}
		\caption{}
		\label{fig:V2 dist}
	\end{subfigure}
	\begin{subfigure}[t]{1 \textwidth}
		\includegraphics[width = .95\linewidth]{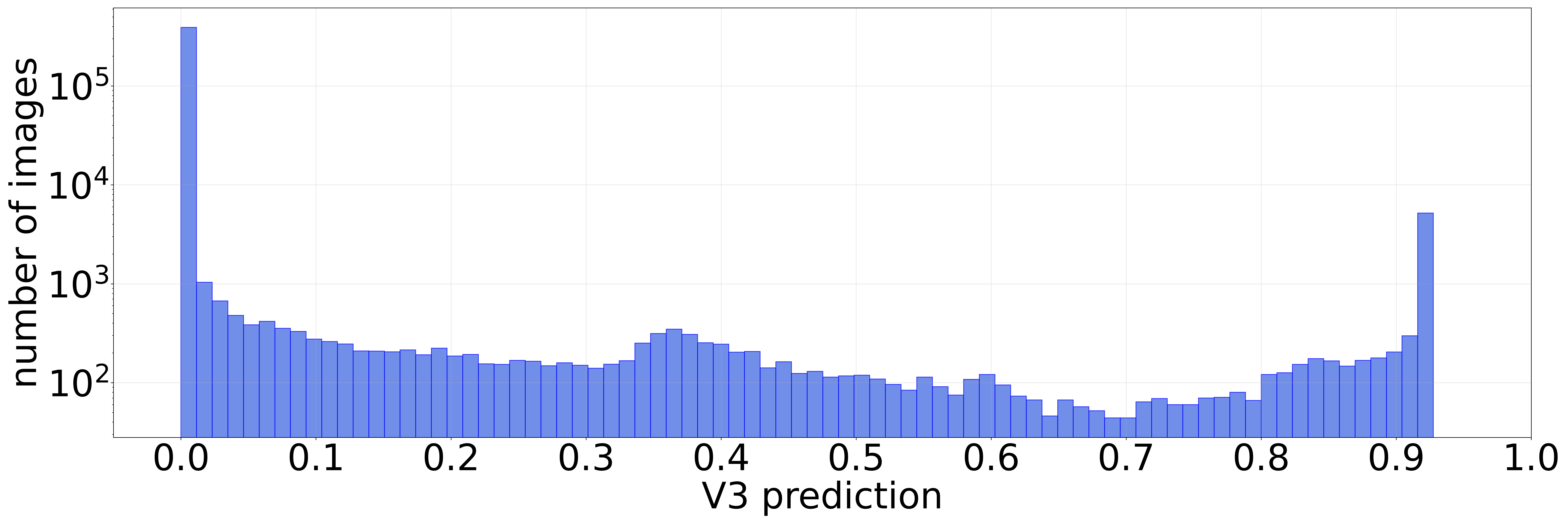}
		\caption{}
		\label{fig:V3 dist}
	\end{subfigure}
	\begin{subfigure}[t]{1 \textwidth}
		\includegraphics[width = .95\linewidth]{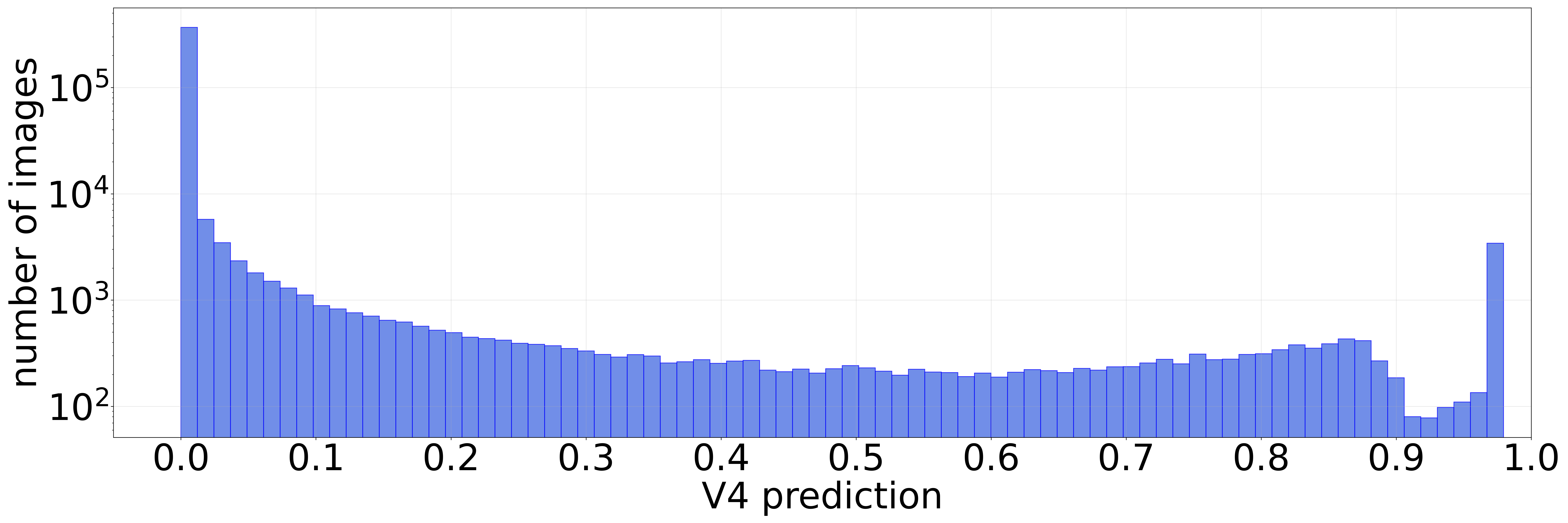}
		\caption{}
		\label{fig:V4 dist}
	\end{subfigure}
	\caption{The distributions of prediction values by V2, V3, and V4, from top to bottom. Note that the y-axis is on a log scale, and that the bin containing the prediction value of 0 is more than 100 times larger than the next largest bin for all three versions.}
	\label{fig:Distributions}
\end{figure}

Only about 15,000 ($\sim4\%$) of the 424,000 images in our dataset meet a threshold of a prediction 0.01 from any of V2, V3, or V4. Thus, the network accomplishes our goal of eliminating the vast majority of images. (See Fig. \ref{fig:Distributions} for detail on the distribution of predictions for all three networks.)


\subsection*{CRATERS FOUND BY OUR ENSEMBLE} 
\begin{figure}[p]
	\centering
	\begin{subfigure}[t]{.48\textwidth}
		\includegraphics[width = .95\linewidth]{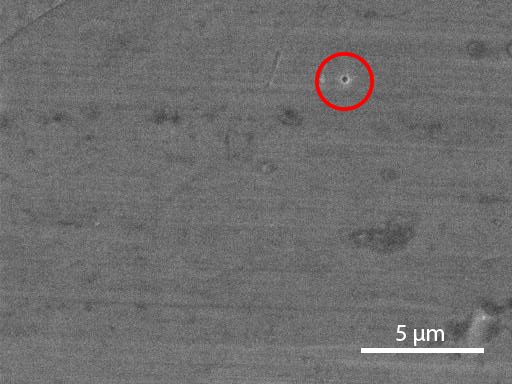} 
		\caption{}
		\label{fig:5600307}
	\end{subfigure}
	\begin{subfigure}[t]{.48\textwidth}
		\includegraphics[width = .95\linewidth]{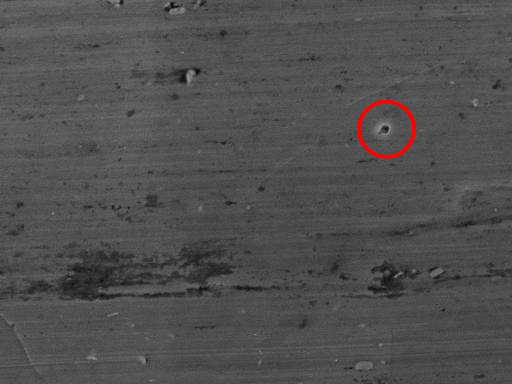} 
		\caption{}
		\label{fig:5658134}
	\end{subfigure}
	\begin{subfigure}[t]{.48\textwidth}
		\includegraphics[width = .95\linewidth]{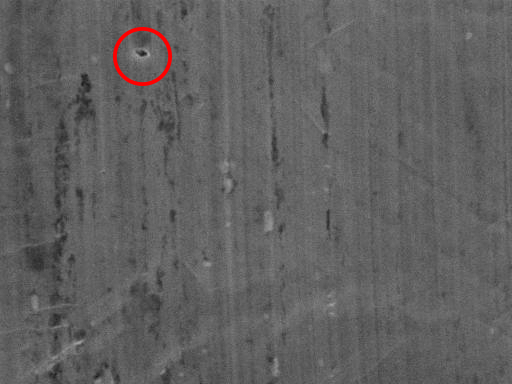}
		\caption{}
		\label{fig:5503346}
	\end{subfigure}
	\begin{subfigure}[t]{.48\textwidth}
		\includegraphics[width = .95\linewidth]{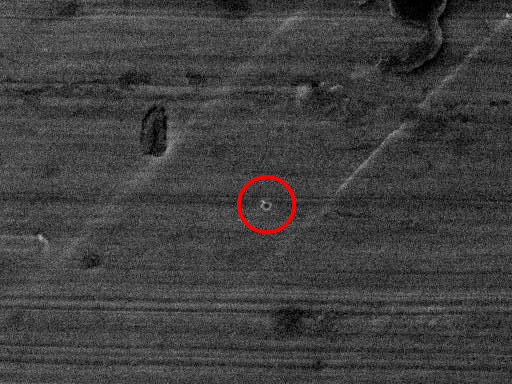}
		\caption{}
		\label{fig:5536887}
	\end{subfigure}

	\caption{Some promising images that may contain craters, as identified by our CNN. (a) and (d) are from the foil I1008W, (b) is from the foil I1126N, and (c) is from the foil I1020W. Prior to the implementation of the network, (b) had one approval and one disapproval, while (c) had no approvals and four disapprovals. (a) and (d) are from a dataset not present on the SAH website. An ``approval" means a volunteer stated that a crater exists in the image, and a ``disapproval" means the volunteer stated no crater was present.}
	\label{fig:promising}
\end{figure}

Approximately 40 promising crater candidates that had previously gone unnoticed were identified by our CNNs. 
Four of these can be seen in Fig. \ref{fig:promising}. 
It will take further examination in order to conclusively determine whether or not these are created by interstellar dust impacts or simply imperfections in the foil. 

Our CNN took about an hour running on an Nvidia GeForce RTX 2060 GPU to generate predictions for the entire dataset. 
This speed is important because it contrasts with the pace of SAH. 
Over the years 2015-2018, the average number of images viewed by volunteers per day was approximately 1,200.
With 424,000 images in the dataset, it took nearly a year ($\frac{424000}{1200} \approx 353$ days) for the SAH volunteers to work through that many images.

Some of the images that contain crater candidates have only been seen once or twice, like Fig. \ref{fig:5658134}, while others have been seen many times and have failed to earn a positive score. 
For example, Fig. \ref{fig:5503346} had been disapproved all 4 times it had viewed.
We would likely never be alerted to images like these, which are prone to being overlooked by volunteers, without an automated approach.

\section*{DISCUSSION} 

\subsection*{A REPLACEMENT OR A COMPLEMENT TO STARDUST@HOME?} 

Deep learning is successful at classification tasks, but has inherent limitations. Whereas a human mind can leverage common sense and relatively few examples to perform well on simple tasks, neural networks lack general knowledge and real understanding, and require a large number of examples to reach proficiency with the same tasks. 

Taking the statistical and practical limitations into account, we opted for a hybrid deployment strategy.  After running the network on all of the images in our dataset, we eliminated all images that the network negatively classified from the SAH website, and left in the ones with predictions greater than 0.01 so that volunteers would have to sort through significantly fewer images without craters.
As we improve the quality of our network, it is possible to gradually reduce our reliance on SAH volunteers, but some form of human engagement will likely always be necessary.


\subsection*{FUTURE IMPROVEMENT} 

As is evident from Table \ref{table:Versions}, there is a rift between our network's true sensitivity and its synthetic sensitivity, even with the augmented dataset used to train V4, which points to an important shortcoming in our algorithm. 
The fact that the synthetic sensitivity is so high means that our network is training effectively: it gets very good at identifying the craters that we feed it.
However, that success doesn't appear to transfer smoothly to the images of true craters in our dataset.
There are two main possibilities.

The first is that our augmentation techniques are inadequate, and we are changing the crater analogs to such an extent that they no longer resemble true craters. 
This is unlikely because we are using established data warping techniques \citep{shorten2019survey} with reasonable augmentation parameters.
Furthermore, we can visually confirm that the augmented craters still resemble genuine craters (Fig. \ref{fig:pasted-150150}).

The second is that the number of analog craters we use to construct our training set is too small.
This is more likely.
Data augmentation, though effective, cannot replace the simple act of collecting more raw data \citep{wong2016understanding, lee2015deeply}.
The only way to reduce the interdependence of training images created through data augmentation is to collect more raw data.

Therefore, we could create more analog craters with the LGG, or re-examine the LGG foils which have already been imaged for new morphologies.
In this way, we can increase the number of analog craters in our dataset by at least a factor of two.
The second way is to let the network find actual Stardust craters.
Even though the true sensitivity isn't perfect, it's still good enough to provide us with new images of true craters and, just as importantly, false positives.
As we gain access to this new data, we can can use this feedback to improve the performance of the network: the network finds some images that it thinks are craters and these images are evaluated both by experts and by volunteers on the SAH website.
These images are then incorporated correspondingly into an improved training set to re-train the network, and this process is iterated with the new version of the CNN.

Data warping-based augmentation is an established technique for improving neural network performance when limited data is available, but there are other approaches to data generation that may be worth exploring. 
An entirely different way to improve our training data in the absence of new raw images is to deploy generative adversarial networks (GANs), which have seen success in recent years as an approach to data augmentation \citep{frid2018gan, shorten2019survey}. 
\cite{perez2017effectiveness} reports GAN-based augmentation alone improves CNN performance about as much as traditional augmentation techniques, but recommends a combination of augmentation techniques that would likely outperform any single technique.


\subsection*{OUTLOOK FOR FURTHER EFFORTS ON STARDUST SAMPLES} 
Aerogel composes 85\% of Stardust's collector area \citep{tsou2003wild}, and preserves particles in a way that aluminum foil does not. 
Thus, it would be useful to have an effective computer vision model for identifying particles in the aerogel section of the collector. 
Here we outline the challenges associated with this application.

The most obvious difference is the dimensionality of the data: when a particle embeds itself in aerogel it creates a track that is imaged in three dimensions. 
On the Stardust@home website, aerogel tracks are viewed as ``movies", in which users view 30 to 40 frames in succession, each successive frame with its focus slightly lower than the last. 
Many tracks cannot be identified by a single frame. 
Much of the literature on volumetric computer vision comes from classification of volumetric medical images, like magnetic resonance imaging (MRI) and computed tomography (CT) scans. 
A number of different approaches to classifying this type of data have been attempted, many of which attempt to adapt 2D networks to this problem: \cite{prasoon2013deep} used three separate 2D CNNs on three orthogonal planes of the volumetric data; \cite{roth2014new} adapted the approach used by \cite{prasoon2013deep} and mapped the three orthogonal planes to the RGB channels of a 2D CNN; \cite{grewal2018radnet} ran a 2D CNN on successive slices in a single plane, but added a long short-term memory (LSTM) layer to their network so previous slices could influence the predictions of later slices. 
Of course, others have attempted to apply 3D CNNs to this type of classification problem, with early attempts by \cite{dou2016automatic} and \cite{payan2015predicting}, and a more recent attempt by \cite{ker2019image} that demonstrated higher performance than previous attempts at reasonable computational cost. 
Though MRI and CT scans produce data of a different form than optical images of aerogel, these results provide optimism that a CNN might be successfully applied to the aerogel section of Stardust's collector. 

\subsection*{APPLICATION TO OTHER MISSIONS} 

Our success with computer vision on the Stardust samples lends credence to the hope that similar CNN methods could be successfully applied to other sample return missions, especially those that collect particulate rather than bulk samples.

The Tanpopo mission, managed by the Japanese Aerospace Exploration Agency (JAXA) and designed to investigate the interplanetary migration of microbes \citep{yamagishi2009tanpopo}, used a sample collection method very similar to the one used by the Stardust mission: aerogel held in place with aluminum foil \citep{tabata2014design}. 
Though it appears little analysis has been done on samples collected by the aluminum foil part of the Tanpopo collector, our research shows that the barrier to beginning this analysis is low, as applying a CNN similar to ours on SEM images of the Tanpopo foils would likely expose many craters, which could then be examined. 
Furthermore, our research lends optimism to an accelerated search for particles trapped in Tanpopo aerogel, as it does for those trapped in Stardust aerogel. 

Stardust's sister mission, Genesis, included collectors that are similar in many respects to Stardust's foils.  Namely, flat surfaces with rare features of interest and many ``contamination" features.  
The main goal of Genesis was to gather data on solar wind \citep{lo1998genesis}, but there has been at least some interest in the analysis of interplanetary dust particles coincidentally caught by the Genesis collectors \citep{graham2004extraction}. 
Also, NASA's Long Duration Exposure Facility (LDEF) performed an experiment to measure the flux of small interplanetary dust particles \citep{singer1985ldef}. 
A re-examination of these surfaces with our CNN could provide an updated calculation.



\section*{CONCLUSION} 
This paper presented the implementation of a CNN to the task of finding craters within SEM images of the aluminum foil portion of the NASA Stardust mission collector. 
The CNN was effective and demonstrated immediate utility. 
We were able to identify crater candidates in the Stardust at Home (SAH) database that had not been noticed before, and we were able to effectively eliminate 96\% of the over 400,000 foils images in the SAH database, leaving SAH volunteers with a much more manageable set of images to look through. 
Even with these successes, our CNN presents clear paths for improvement. 
Particularly, the collection of more analog and true craters would increase the size of our training set and improve the effectiveness of our network. 
Another way to address the small amount of craters we currently have is to use techniques like GAN-based augmentation.
This CNN might be a practical and convenient tool to future analyses of Stardust foils, and our CNN could be applied to sample return missions with similar collectors, such as Genesis and Tanpopo. 
The success of deep learning methods on Stardust foils further gives optimism that deep learning methods can be applied to Stardust's aerogel collector.

\section*{ACKNOWLEDGEMENTS}
Computational work at the Advanced Light Source and Molecular Foundry at Lawrence Berkeley National Laboratory was supported by the U.S. Department of Energy under Contract No. DE-AC02- 05CH11231.
This work was supported by NASA grant NNX16AI15G.

We specifically thank David Shapiro for providing access to computing resources, allowing us to train the CNN.

\medskip

\renewcommand\refname{REFERENCES}

\bibliographystyle{MAPS}
\addcontentsline{toc}{section}{\refname}\bibliography{main_nc.bib}

\begin{thebibliography}{}

\bibitem[\protect\astroncite{Abadi et~al.}{2015}]{tensorflow2015-whitepaper}
Abadi, M., Agarwal, A., Barham, P., Brevdo, E., Chen, Z., Citro, C., Corrado,
  G.~S., Davis, A., Dean, J., Devin, M., Ghemawat, S., Goodfellow, I., Harp,
  A., Irving, G., Isard, M., Jia, Y., Jozefowicz, R., Kaiser, L., Kudlur, M.,
  Levenberg, J., Man\'{e}, D., Monga, R., Moore, S., Murray, D., Olah, C.,
  Schuster, M., Shlens, J., Steiner, B., Sutskever, I., Talwar, K., Tucker, P.,
  Vanhoucke, V., Vasudevan, V., Vi\'{e}gas, F., Vinyals, O., Warden, P.,
  Wattenberg, M., Wicke, M., Yu, Y., and Zheng, X. 2015.
\newblock {TensorFlow}: Large-scale machine learning on heterogeneous systems.
\newblock Software available from tensorflow.org.

\bibitem[\protect\astroncite{Bechtel et~al.}{2014}]{bechtel2014stardust}
Bechtel, H.~A., Flynn, G.~J., Allen, C., Anderson, D., Ansari, A., Bajt, S.,
  Bastien, R.~K., Bassim, N., Borg, J., Brenker, F.~E., et~al. 2014.
\newblock Stardust interstellar preliminary examination iii: Infrared
  spectroscopic analysis of interstellar dust candidates.
\newblock {\em Meteoritics \& Planetary Science} 49:1548--1561.

\bibitem[\protect\astroncite{Brenker et~al.}{2014}]{brenker2014stardust}
Brenker, F.~E., Westphal, A.~J., Vincze, L., Burghammer, M., Schmitz, S.,
  Schoonjans, T., Silversmit, G., Vekemans, B., Allen, C., Anderson, D., et~al.
  2014.
\newblock Stardust interstellar preliminary examination v: Xrf analyses of
  interstellar dust candidates at esrf id 13.
\newblock {\em Meteoritics \& Planetary Science} 49:1594--1611.

\bibitem[\protect\astroncite{Brownlee}{2014}]{brownlee2014review}
Brownlee, D. 2014.
\newblock The stardust mission: Analyzing samples from the edge of the solar
  system.
\newblock {\em Annual Review of Earth and Planetary Sciences} 42:179--205.

\bibitem[\protect\astroncite{Brownlee et~al.}{2006}]{brownlee2006comet}
Brownlee, D., Tsou, P., Al{\'e}on, J., Alexander, C.~M., Araki, T., Bajt, S.,
  Baratta, G.~A., Bastien, R., Bland, P., Bleuet, P., et~al. 2006.
\newblock Comet 81p/wild 2 under a microscope.
\newblock {\em science} 314:1711--1716.

\bibitem[\protect\astroncite{Brownlee et~al.}{2004}]{brownlee2004surface}
Brownlee, D.~E., Horz, F., Newburn, R.~L., Zolensky, M., Duxbury, T.~C.,
  Sandford, S., Sekanina, Z., Tsou, P., Hanner, M.~S., Clark, B.~C., et~al.
  2004.
\newblock Surface of young jupiter family comet 81p/wild 2: View from the
  stardust spacecraft.
\newblock {\em Science} 304:1764--1769.

\bibitem[\protect\astroncite{Burchell et~al.}{1999}]{burchell1999hypervelocity}
Burchell, M.~J., Cole, M.~J., McDonnell, J., and Zarnecki, J.~C. 1999.
\newblock Hypervelocity impact studies using the 2 mv van de graaff accelerator
  and two-stage light gas gun of the university of kent at canterbury.
\newblock {\em Measurement Science and Technology} 10:41.

\bibitem[\protect\astroncite{Burchell et~al.}{2012}]{burchell2012experimental}
Burchell, M.~J., Cole, M.~J., Price, M.~C., and Kearsley, A.~T. 2012.
\newblock Experimental investigation of impacts by solar cell secondary ejecta
  on silica aerogel and aluminum foil: Implications for the stardust
  interstellar dust collector.
\newblock {\em Meteoritics \& Planetary Science} 47:671--683.

\bibitem[\protect\astroncite{Butterworth
  et~al.}{2014}]{butterworth2014stardust}
Butterworth, A.~L., Westphal, A.~J., Tyliszczak, T., Gainsforth, Z., Stodolna,
  J., Frank, D.~R., Allen, C., Anderson, D., Ansari, A., Bajt, S., et~al. 2014.
\newblock Stardust interstellar preliminary examination iv: Scanning
  transmission x-ray microscopy analyses of impact features in the stardust
  interstellar dust collector.
\newblock {\em Meteoritics \& Planetary Science} 49:1562--1593.

\bibitem[\protect\astroncite{Caruana et~al.}{2004}]{caruana2004ensemble}
Caruana, R., Niculescu-Mizil, A., Crew, G., and Ksikes, A. 2004.
\newblock Ensemble selection from libraries of models.
\newblock {\em In} Proceedings of the twenty-first international conference on
  Machine learning, p.~18.

\bibitem[\protect\astroncite{Chollet et~al.}{2015}]{chollet2015keras}
Chollet, F. et~al. 2015.
\newblock Keras, https://keras.io.

\bibitem[\protect\astroncite{De~Gregorio et~al.}{2018}]{de2018fast}
De~Gregorio, B., Brintlinger, T., and Stroud, R. 2018.
\newblock Fast automated identification of dust impact craters in aluminum
  foils from the stardust collection.
\newblock {\em In} Lunar and Planetary Science Conference, volume~49.

\bibitem[\protect\astroncite{Dietterich}{2000}]{dietterich2000ensemble}
Dietterich, T.~G. 2000.
\newblock Ensemble methods in machine learning.
\newblock {\em In} International workshop on multiple classifier systems, pp.
  1--15. Springer.

\bibitem[\protect\astroncite{Dou et~al.}{2016}]{dou2016automatic}
Dou, Q., Chen, H., Yu, L., Zhao, L., Qin, J., Wang, D., Mok, V.~C., Shi, L.,
  and Heng, P.-A. 2016.
\newblock Automatic detection of cerebral microbleeds from mr images via 3d
  convolutional neural networks.
\newblock {\em IEEE transactions on medical imaging} 35:1182--1195.

\bibitem[\protect\astroncite{Dozat}{2016}]{Dozat2016IncorporatingNM}
Dozat, T. 2016.
\newblock Incorporating nesterov momentum into adam.

\bibitem[\protect\astroncite{Elsila et~al.}{2009}]{elsila2009cometary}
Elsila, J.~E., Glavin, D.~P., and Dworkin, J.~P. 2009.
\newblock Cometary glycine detected in samples returned by stardust.
\newblock {\em Meteoritics \& Planetary Science} 44:1323--1330.

\bibitem[\protect\astroncite{Flynn et~al.}{2006}]{flynn2006elemental}
Flynn, G.~J., Bleuet, P., Borg, J., Bradley, J.~P., Brenker, F.~E., Brennan,
  S., Bridges, J., Brownlee, D.~E., Bullock, E.~S., Burghammer, M., et~al.
  2006.
\newblock Elemental compositions of comet 81p/wild 2 samples collected by
  stardust.
\newblock {\em Science} 314:1731--1735.

\bibitem[\protect\astroncite{Flynn et~al.}{2014}]{flynn2014stardust}
Flynn, G.~J., Sutton, S.~R., Lai, B., Wirick, S., Allen, C., Anderson, D.,
  Ansari, A., Bajt, S., Bastien, R.~K., Bassim, N., et~al. 2014.
\newblock Stardust interstellar preliminary examination vii: Synchrotron x-ray
  fluorescence analysis of six stardust interstellar candidates measured with
  the advanced photon source 2-id-d microprobe.
\newblock {\em Meteoritics \& Planetary Science} 49:1626--1644.

\bibitem[\protect\astroncite{Frid-Adar et~al.}{2018}]{frid2018gan}
Frid-Adar, M., Diamant, I., Klang, E., Amitai, M., Goldberger, J., and
  Greenspan, H. 2018.
\newblock Gan-based synthetic medical image augmentation for increased cnn
  performance in liver lesion classification.
\newblock {\em Neurocomputing} 321:321--331.

\bibitem[\protect\astroncite{Gainsforth et~al.}{2014}]{gainsforth2014stardust}
Gainsforth, Z., Brenker, F.~E., Simionovici, A.~S., Schmitz, S., Burghammer,
  M., Butterworth, A.~L., Cloetens, P., Lemelle, L., Tresserras, J.-A.~S.,
  Schoonjans, T., et~al. 2014.
\newblock Stardust interstellar preliminary examination viii: Identification of
  crystalline material in two interstellar candidates.
\newblock {\em Meteoritics \& Planetary Science} 49:1645--1665.

\bibitem[\protect\astroncite{Gainsforth et~al.}{2019}]{gainsforth2019fine}
Gainsforth, Z., Westphal, A.~J., Butterworth, A.~L., Jilly-Rehak, C.~E.,
  Brownlee, D.~E., Joswiak, D., Ogliore, R.~C., Zolensky, M.~E., Bechtel,
  H.~A., Ebel, D.~S., et~al. 2019.
\newblock Fine-grained material associated with a large sulfide returned from
  comet 81p/wild 2.
\newblock {\em Meteoritics \& planetary science} 54:1069--1091.

\bibitem[\protect\astroncite{Giusti et~al.}{2013}]{giusti2013fast}
Giusti, A., Cire{\c{s}}an, D.~C., Masci, J., Gambardella, L.~M., and
  Schmidhuber, J. 2013.
\newblock Fast image scanning with deep max-pooling convolutional neural
  networks.
\newblock {\em In} 2013 IEEE International Conference on Image Processing, pp.
  4034--4038. IEEE.

\bibitem[\protect\astroncite{Graham et~al.}{2004}]{graham2004extraction}
Graham, G.~A., Kearsley, A.~T., Butterworth, A., Bland, P.~A., Burchell, M.~J.,
  McPhail, D.~S., Chater, R., Grady, M.~M., and Wright, I.~P. 2004.
\newblock Extraction and microanalysis of cosmic dust captured during sample
  return missions: laboratory simulations.
\newblock {\em Advances in Space Research} 34:2292--2298.

\bibitem[\protect\astroncite{Grewal et~al.}{2018}]{grewal2018radnet}
Grewal, M., Srivastava, M.~M., Kumar, P., and Varadarajan, S. 2018.
\newblock Radnet: Radiologist level accuracy using deep learning for hemorrhage
  detection in ct scans.
\newblock {\em In} 2018 IEEE 15th International Symposium on Biomedical Imaging
  (ISBI 2018), pp. 281--284. IEEE.

\bibitem[\protect\astroncite{He et~al.}{2015}]{he_delving_2015}
He, K., Zhang, X., Ren, S., and Sun, J. 2015.
\newblock Delving {Deep} into {Rectifiers}: {Surpassing} {Human}-{Level}
  {Performance} on {ImageNet} {Classification}.
\newblock {\em In} 2015 {IEEE} {International} {Conference} on {Computer}
  {Vision} ({ICCV}), pp. 1026--1034.
\newblock ISSN: 2380-7504.

\bibitem[\protect\astroncite{H{\"o}rz et~al.}{2006}]{horz2006impacts}
H{\"o}rz, F., Bastien, R., Borg, J., Bradley, J.~P., Bridges, J.~C., Brownlee,
  D.~E., Burchell, M.~J., Chi, M., Cintala, M.~J., Dai, Z.~R., Djouadi, Z.,
  Dominguez, G., Economou, T.~E., Fairey, S. A.~J., Floss, C., Franchi, I.~A.,
  Graham, G.~A., Green, S.~F., Heck, P., Hoppe, P., Huth, J., Ishii, H.,
  Kearsley, A.~T., Kissel, J., Leitner, J., Leroux, H., Marhas, K., Messenger,
  K., Schwandt, C.~S., See, T.~H., Snead, C., Stadermann~I, F.~J., Stephan, T.,
  Stroud, R., Teslich, N., Trigo-Rodriguez, J.~M., Tuzzolino, A.~J., Troadec,
  D., Tsou, P., Warren, J., Westphal, A.~J., Wozniakiewicz, P., Wright, I., and
  Zinner, E. 2006.
\newblock {Impact Features on Stardust: Implications for Comet 81P/Wild 2
  Dust}.
\newblock {\em Science} 314:1716--1719.

\bibitem[\protect\astroncite{Huang et~al.}{2017}]{huang2017snapshot}
Huang, G., Li, Y., Pleiss, G., Liu, Z., Hopcroft, J.~E., and Weinberger, K.~Q.
  2017.
\newblock Snapshot ensembles: Train 1, get m for free.
\newblock {\em arXiv preprint arXiv:1704.00109} .

\bibitem[\protect\astroncite{Hunter}{2007}]{Hunter:2007}
Hunter, J.~D. 2007.
\newblock Matplotlib: A 2d graphics environment.
\newblock {\em Computing in Science \& Engineering} 9:90--95.

\bibitem[\protect\astroncite{Joswiak et~al.}{2012}]{joswiak2012comprehensive}
Joswiak, D.~J., Brownlee, D.~E., Matrajt, G., Westphal, A.~J., Snead, C.~J.,
  and Gainsforth, Z. 2012.
\newblock Comprehensive examination of large mineral and rock fragments in
  stardust tracks: Mineralogy, analogous extraterrestrial materials, and source
  regions.
\newblock {\em Meteoritics \& Planetary Science} 47:471--524.

\bibitem[\protect\astroncite{Kearsley et~al.}{2008}]{kearsley2008dust}
Kearsley, A.~T., Borg, J., Graham, G.~A., Burchell, M.~J., Cole, M.~J., Leroux,
  H., Bridges, J.~C., H{\"o}rz, F., Wozniakiewicz, P.~J., Bland, P.~A., et~al.
  2008.
\newblock Dust from comet wild 2: Interpreting particle size, shape, structure,
  and composition from impact features on the stardust aluminum foils.
\newblock {\em Meteoritics \& Planetary Science} 43:41--73.

\bibitem[\protect\astroncite{Ker et~al.}{2019}]{ker2019image}
Ker, J., Singh, S.~P., Bai, Y., Rao, J., Lim, T., and Wang, L. 2019.
\newblock Image thresholding improves 3-dimensional convolutional neural
  network diagnosis of different acute brain hemorrhages on computed tomography
  scans.
\newblock {\em Sensors} 19:2167.

\bibitem[\protect\astroncite{Kluyver et~al.}{2016}]{kluyver2016jupyter}
Kluyver, T., Ragan-Kelley, B., P{\'e}rez, F., Granger, B.~E., Bussonnier, M.,
  Frederic, J., Kelley, K., Hamrick, J.~B., Grout, J., Corlay, S., et~al. 2016.
\newblock Jupyter notebooks-a publishing format for reproducible computational
  workflows.
\newblock {\em In} ELPUB, pp. 87--90.

\bibitem[\protect\astroncite{Krizhevsky
  et~al.}{2012}]{krizhevsky_imagenet_2012}
Krizhevsky, A., Sutskever, I., and Hinton, G.~E. 2012.
\newblock {ImageNet} {Classification} with {Deep} {Convolutional} {Neural}
  {Networks}, pp. 1097--1105.
\newblock {\em In} F. Pereira, C.~J.~C. Burges, L. Bottou, and K.~Q. Weinberger
  (eds.), Advances in {Neural} {Information} {Processing} {Systems} 25. Curran
  Associates, Inc.

\bibitem[\protect\astroncite{Krogh and Hertz}{1992}]{krogh1992simple}
Krogh, A. and Hertz, J.~A. 1992.
\newblock A simple weight decay can improve generalization.
\newblock {\em In} Advances in neural information processing systems, pp.
  950--957.

\bibitem[\protect\astroncite{Landgraf et~al.}{1999}]{landgraf1999prediction}
Landgraf, M., M{\"u}ller, M., and Gr{\"u}n, E. 1999.
\newblock Prediction of the in-situ dust measurements of the stardust mission
  to comet 81p/wild 2.
\newblock {\em Planetary and Space Science} 47:1029--1050.

\bibitem[\protect\astroncite{Lee et~al.}{2015}]{lee2015deeply}
Lee, C.-Y., Xie, S., Gallagher, P., Zhang, Z., and Tu, Z. 2015.
\newblock Deeply-supervised nets.
\newblock {\em In} Artificial intelligence and statistics, pp. 562--570.

\bibitem[\protect\astroncite{Leitner et~al.}{2008}]{Leitner}
Leitner, J., Stephan, T., Kearsley, A.~T., H{\"o}rz, F., Flynn, G.~J., and
  Sandford, S.~A. 2008.
\newblock Tof‐sims analysis of crater residues from wild 2 cometary particles
  on stardust aluminum foil.
\newblock {\em Meteoritics \& Planetary Science} 43:161--185.

\bibitem[\protect\astroncite{LeNail}{2019}]{lenail2019nn}
LeNail, A. 2019.
\newblock Nn-svg: Publication-ready neural network architecture schematics.
\newblock {\em Journal of Open Source Software} 4:747.

\bibitem[\protect\astroncite{Leroux et~al.}{2008}]{leroux2008transmission}
Leroux, H., STroud, R.~M., Dai, Z.~R., Graham, G.~A., Troadec, D., Bradley,
  J.~P., Teslich, N., Borg, J., Kearsley, A.~T., and H{\"o}rz, F. 2008.
\newblock Transmission electron microscopy of cometary residues from
  micron-sized craters in the stardust al foils.
\newblock {\em Meteoritics \& Planetary Science} 43:143--160.

\bibitem[\protect\astroncite{Lin et~al.}{2013}]{lin2013network}
Lin, M., Chen, Q., and Yan, S. 2013.
\newblock Network in network.
\newblock {\em arXiv preprint arXiv:1312.4400} .

\bibitem[\protect\astroncite{Lo et~al.}{1998}]{lo1998genesis}
Lo, M., Williams, B., Bollman, W., Han, D., Hahn, Y., Bell, J., Hirst, E.,
  Corwin, R., Hong, P., and Howell, K. 1998.
\newblock Genesis mission design.
\newblock {\em In} AIAA/AAS Astrodynamics Specialist Conference and Exhibit, p.
  4468.

\bibitem[\protect\astroncite{McKeegan et~al.}{2006}]{mckeegan2006isotopic}
McKeegan, K.~D., Al{\'e}on, J., Bradley, J., Brownlee, D., Busemann, H.,
  Butterworth, A., Chaussidon, M., Fallon, S., Floss, C., Gilmour, J., et~al.
  2006.
\newblock Isotopic compositions of cometary matter returned by stardust.
\newblock {\em Science} 314:1724--1728.

\bibitem[\protect\astroncite{Ogliore et~al.}{2012a}]{ogliore2012incorporation}
Ogliore, R., Huss, G., Nagashima, K., Butterworth, A., Gainsforth, Z.,
  Stodolna, J., Westphal, A., Joswiak, D., and Tyliszczak, T. 2012a.
\newblock Incorporation of a late-forming chondrule into comet wild 2.
\newblock {\em The Astrophysical Journal Letters} 745:L19.

\bibitem[\protect\astroncite{Ogliore et~al.}{2012b}]{ogliore2012automated}
Ogliore, R.~C., Floss, C., Stadermann, F.~J., Kearsley, A., Leitner, J.,
  Stroud, R.~M., and Westphal, A.~J. 2012b.
\newblock Automated searching of stardust interstellar foils.
\newblock {\em Meteoritics \& Planetary Science} 47:729--736.

\bibitem[\protect\astroncite{Payan and Montana}{2015}]{payan2015predicting}
Payan, A. and Montana, G. 2015.
\newblock Predicting alzheimer's disease: a neuroimaging study with 3d
  convolutional neural networks.
\newblock {\em arXiv preprint arXiv:1502.02506} .

\bibitem[\protect\astroncite{Pedregosa et~al.}{2011}]{scikit-learn}
Pedregosa, F., Varoquaux, G., Gramfort, A., Michel, V., Thirion, B., Grisel,
  O., Blondel, M., Prettenhofer, P., Weiss, R., Dubourg, V., Vanderplas, J.,
  Passos, A., Cournapeau, D., Brucher, M., Perrot, M., and Duchesnay, E. 2011.
\newblock Scikit-learn: Machine learning in {P}ython.
\newblock {\em Journal of Machine Learning Research} 12:2825--2830.

\bibitem[\protect\astroncite{Perez and Wang}{2017}]{perez2017effectiveness}
Perez, L. and Wang, J. 2017.
\newblock The effectiveness of data augmentation in image classification using
  deep learning.
\newblock {\em arXiv preprint arXiv:1712.04621} .

\bibitem[\protect\astroncite{Prasoon et~al.}{2013}]{prasoon2013deep}
Prasoon, A., Petersen, K., Igel, C., Lauze, F., Dam, E., and Nielsen, M. 2013.
\newblock Deep feature learning for knee cartilage segmentation using a
  triplanar convolutional neural network.
\newblock {\em In} International conference on medical image computing and
  computer-assisted intervention, pp. 246--253. Springer.

\bibitem[\protect\astroncite{Price et~al.}{2010}]{price2010}
Price, M., Kearsley, A., Burchell, M., Hörz, F., Borg, J., Bridges, J., Cole,
  M., Floss, C., Graham, G., Green, S., Hoppe, P., Leroux, H., Marhas, K.,
  Park, N., Stroud, R., FJ, S., and PJ, W. 2010.
\newblock Comet 81p/wild 2: The size distribution of finer (sub-10 micron) dust
  collected by the stardust spacecraft.
\newblock {\em Meteoritics \& Planetary Science} 45:1409--1428.

\bibitem[\protect\astroncite{Roth et~al.}{2014}]{roth2014new}
Roth, H.~R., Lu, L., Seff, A., Cherry, K.~M., Hoffman, J., Wang, S., Liu, J.,
  Turkbey, E., and Summers, R.~M. 2014.
\newblock A new 2.5 d representation for lymph node detection using random sets
  of deep convolutional neural network observations.
\newblock {\em In} International conference on medical image computing and
  computer-assisted intervention, pp. 520--527. Springer.

\bibitem[\protect\astroncite{Sandford et~al.}{2006}]{sandford2006organics}
Sandford, S.~A., Al{\'e}on, J., Alexander, C.~M., Araki, T., Bajt, S., Baratta,
  G.~A., Borg, J., Bradley, J.~P., Brownlee, D.~E., Brucato, J.~R., et~al.
  2006.
\newblock Organics captured from comet 81p/wild 2 by the stardust spacecraft.
\newblock {\em Science} 314:1720--1724.

\bibitem[\protect\astroncite{Shorten and
  Khoshgoftaar}{2019}]{shorten2019survey}
Shorten, C. and Khoshgoftaar, T.~M. 2019.
\newblock A survey on image data augmentation for deep learning.
\newblock {\em Journal of Big Data} 6:60.

\bibitem[\protect\astroncite{Simon et~al.}{2008}]{simon2008refractory}
Simon, S.~B., Joswiak, D., Ishii, H., Bradley, J.~P., Chi, M., Grossman, L.,
  Aleon, J., Brownlee, D., Fallon, S., Hutcheon, I.~D., et~al. 2008.
\newblock A refractory inclusion returned by stardust from comet 81p/wild 2.
\newblock {\em Meteoritics \& Planetary Science} 43:1861--1877.

\bibitem[\protect\astroncite{Simonyan and Zisserman}{2014}]{simonyan2014very}
Simonyan, K. and Zisserman, A. 2014.
\newblock Very deep convolutional networks for large-scale image recognition.
\newblock {\em arXiv preprint arXiv:1409.1556} .

\bibitem[\protect\astroncite{Singer et~al.}{1985}]{singer1985ldef}
Singer, S., Stanley, J., and Kassel, P. 1985.
\newblock The ldef interplanetary dust experiment.
\newblock {\em In} International Astronomical Union Colloquium, volume~85, pp.
  117--120. Cambridge University Press.

\bibitem[\protect\astroncite{Srivastava et~al.}{2014}]{JMLR:v15:srivastava14a}
Srivastava, N., Hinton, G., Krizhevsky, A., Sutskever, I., and Salakhutdinov,
  R. 2014.
\newblock Dropout: A simple way to prevent neural networks from overfitting.
\newblock {\em Journal of Machine Learning Research} 15:1929--1958.

\bibitem[\protect\astroncite{Stadermann et~al.}{2009}]{stadermann2009use}
Stadermann, F.~J., Floss, C., Bose, M., and Lea, A.~S. 2009.
\newblock The use of auger spectroscopy for the in situ elemental
  characterization of sub-micrometer presolar grains.
\newblock {\em Meteoritics \& Planetary Science} 44:1033--1049.

\bibitem[\protect\astroncite{Stadermann et~al.}{2008}]{stadermann2008stardust}
Stadermann, F.~J., Hoppe, P., Floss, C., Heck, P.~R., H{\"o}rz, F., Huth, J.,
  Kearsley, A.~T., Leitner, J., Marhas, K.~K., Mckeegan, K.~D., et~al. 2008.
\newblock Stardust in stardust—the c, n, and o isotopic compositions of wild
  2 cometary matter in al foil impacts.
\newblock {\em Meteoritics \& Planetary Science} 43:299--313.

\bibitem[\protect\astroncite{Stroud et~al.}{2010}]{stroud2010structure}
Stroud, R., Koch, I., Bassim, N., Piccard, Y., and Nittler, L. 2010.
\newblock Structure and composition of comet wild 2 residues in sub-micron to
  micron-sized craters.
\newblock {\em LPI} p. 1792.

\bibitem[\protect\astroncite{Stroud et~al.}{2014}]{stroud2014stardust}
Stroud, R.~M., Allen, C., Ansari, A., Anderson, D., Bajt, S., Bassim, N.,
  Bastien, R.~S., Bechtel, H.~A., Borg, J., Brenker, F.~E., et~al. 2014.
\newblock Stardust interstellar preliminary examination xi: Identification and
  elemental analysis of impact craters on al foils from the stardust
  interstellar dust collector.
\newblock {\em Meteoritics \& Planetary Science} 49:1698--1719.

\bibitem[\protect\astroncite{Tabata et~al.}{2014}]{tabata2014design}
Tabata, M., Imai, E., Yano, H., Hashimoto, H., Kawai, H., Kawaguchi, Y.,
  Kobayashi, K., Mita, H., Okudaira, K., Sasaki, S., et~al. 2014.
\newblock Design of a silica-aerogel-based cosmic dust collector for the
  tanpopo mission aboard the international space station.
\newblock {\em Transactions of the Japan Society for Aeronautical and Space
  Sciences, Aerospace Technology Japan} 12:Pk\_29--Pk\_34.

\bibitem[\protect\astroncite{{The GIMP Development Team,
  https://www.gimp.org}}{2019}]{gimp}
{The GIMP Development Team, https://www.gimp.org} 2019.
\newblock Gimp.

\bibitem[\protect\astroncite{{The HDF Group}}{1997}]{hdf5}
{The HDF Group} 1997.
\newblock {Hierarchical Data Format, version 5}.
\newblock http://www.hdfgroup.org/HDF5/.

\bibitem[\protect\astroncite{Tompson et~al.}{2015}]{Tompson2015EfficientOL}
Tompson, J., Goroshin, R., Jain, A., LeCun, Y., and Bregler, C. 2015.
\newblock Efficient object localization using convolutional networks.
\newblock {\em 2015 IEEE Conference on Computer Vision and Pattern Recognition
  (CVPR)} pp. 648--656.

\bibitem[\protect\astroncite{Tsou et~al.}{2003}]{tsou2003wild}
Tsou, P., Brownlee, D., Sandford, S., H{\"o}rz, F., and Zolensky, M. 2003.
\newblock Wild 2 and interstellar sample collection and earth return.
\newblock {\em Journal of Geophysical Research: Planets} 108.

\bibitem[\protect\astroncite{{van der Walt} et~al.}{2011}]{5725236}
{van der Walt}, S., {Colbert}, S.~C., and {Varoquaux}, G. 2011.
\newblock The numpy array: A structure for efficient numerical computation.
\newblock {\em Computing in Science Engineering} 13:22--30.

\bibitem[\protect\astroncite{van~der Walt et~al.}{2014}]{scikit-image}
van~der Walt, S., {S}ch\"onberger, J.~L., {Nunez-Iglesias}, J., {B}oulogne, F.,
  {W}arner, J.~D., {Y}ager, N., {G}ouillart, E., {Y}u, T., and the scikit-image
  contributors 2014.
\newblock scikit-image: image processing in {P}ython.
\newblock {\em PeerJ} 2:e453.

\bibitem[\protect\astroncite{Westphal et~al.}{2014a}]{westphal2014stardust}
Westphal, A.~J., Anderson, D., Butterworth, A.~L., Frank, D.~R., Lettieri, R.,
  Marchant, W., Von~Korff, J., Zevin, D., Ardizzone, A., Campanile, A., et~al.
  2014a.
\newblock Stardust interstellar preliminary examination i: identification of
  tracks in aerogel.
\newblock {\em Meteoritics \& Planetary Science} 49:1509--1521.

\bibitem[\protect\astroncite{Westphal et~al.}{2014b}]{westphal2014final}
Westphal, A.~J., Bechtel, H.~A., Brenker, F.~E., Butterworth, A.~L., Flynn, G.,
  Frank, D.~R., Gainsforth, Z., Hillier, J.~K., Postberg, F., Simionovici,
  A.~S., et~al. 2014b.
\newblock Final reports of the stardust interstellar preliminary examination.
\newblock {\em Meteoritics \& Planetary Science} 49:1720--1733.

\bibitem[\protect\astroncite{Westphal et~al.}{2005}]{westphal2005stardust}
Westphal, A.~J., Butterworth, A.~L., Snead, C.~J., Craig, N., Anderson, D.,
  Jones, S.~M., Brownlee, D.~E., Farnsworth, R., and Zolensky, M.~E. 2005.
\newblock Stardust@ home: a massively distributed public search for
  interstellar dust in the stardust interstellar dust collector.

\bibitem[\protect\astroncite{Westphal et~al.}{2014c}]{westphal2014evidence}
Westphal, A.~J., Stroud, R.~M., Bechtel, H.~A., Brenker, F.~E., Butterworth,
  A.~L., Flynn, G.~J., Frank, D.~R., Gainsforth, Z., Hillier, J.~K., Postberg,
  F., et~al. 2014c.
\newblock Evidence for interstellar origin of seven dust particles collected by
  the stardust spacecraft.
\newblock {\em science} 345:786--791.

\bibitem[\protect\astroncite{Wong et~al.}{2016}]{wong2016understanding}
Wong, S.~C., Gatt, A., Stamatescu, V., and McDonnell, M.~D. 2016.
\newblock Understanding data augmentation for classification: when to warp?
\newblock {\em In} 2016 international conference on digital image computing:
  techniques and applications (DICTA), pp. 1--6. IEEE.

\bibitem[\protect\astroncite{Wozniakiewicz
  et~al.}{2012}]{wozniakiewicz2012stardust}
Wozniakiewicz, P.~J., Ishii, H.~A., Kearsley, A.~T., Burchell, M.~J., Bradley,
  J.~P., Price, M.~C., Teslich, N., Lee, M.~R., and Cole, M.~J. 2012.
\newblock Stardust impact analogs: Resolving pre-and postimpact mineralogy in
  stardust al foils.
\newblock {\em Meteoritics \& Planetary Science} 47:708--728.

\bibitem[\protect\astroncite{Yamagishi et~al.}{2009}]{yamagishi2009tanpopo}
Yamagishi, A., Yano, H., Okudaira, K., Kobayashi, K., Yokobori, S.-I., Tabata,
  M., Kawai, H., Yamashita, M., Hashimoto, H., Naraoka, H., et~al. 2009.
\newblock Tanpopo: Astrobiology exposure and micrometeoroid capture
  experiments.
\newblock {\em Transactions of the Japan Society for Aeronautical and Space
  Sciences, Space Technology Japan} 7:Tk\_49--Tk\_55.

\bibitem[\protect\astroncite{Yu et~al.}{2017}]{yu2017deep}
Yu, X., Wu, X., Luo, C., and Ren, P. 2017.
\newblock Deep learning in remote sensing scene classification: a data
  augmentation enhanced convolutional neural network framework.
\newblock {\em GIScience \& Remote Sensing} 54:741--758.

\bibitem[\protect\astroncite{Zolensky et~al.}{2006}]{zolensky2006mineralogy}
Zolensky, M.~E., Zega, T.~J., Yano, H., Wirick, S., Westphal, A.~J., Weisberg,
  M.~K., Weber, I., Warren, J.~L., Velbel, M.~A., Tsuchiyama, A., et~al. 2006.
\newblock Mineralogy and petrology of comet 81p/wild 2 nucleus samples.
\newblock {\em Science} 314:1735--1739.

\end{thebibliography}

\begin{table}[p]
    \centering
    \begin{tabular}{cc}
        \textbf{Layer} & \textbf{Units} \\ \hline
        Conv2D         & 64             \\ 
        Conv2D         & 64             \\ 
        MaxPool        & -              \\ 
        Conv2D         & 64             \\ 
        Conv2D         & 64             \\ 
        MaxPool        & -              \\ 
        Conv2D         & 64             \\ 
        Conv2D         & 64             \\ 
        MaxPool        & -              \\ 
        Conv2D         & 32             \\ 
        MaxPool        & -              \\ 
        Conv2D         & 32             \\ 
        Global MaxPool & -              \\ 
        Dense          & 32             \\ 
        Dense          & 32             \\ 
        Dense          & 16             \\ 
        Dense          & 16             \\ 
        Dense          & 16             \\ 
        Dense          & 16             \\ 
        Dense          & 1              \\ 
    \end{tabular}
    \caption{Simplified tablular representation of the network. See the Network Architecture section for a more detailed description. The ``units" column here represents the final dimension of the output space of the layer. That is, it is equal to the number of filters in a convolutional layer, and the number of output nodes in a densely connected layer.}
    \label{table:network}
\end{table}

\begin{table}[p]
	\centering
	\begin{tabular}{c|ccc}
		                                    & \textbf{V2} & \textbf{V3} & \textbf{V4} \\ \hline
		Number of 150x150 Training Images   & 6,000        & 20,000      & 20,000      \\
		Number of Full-Size Training Images & 2,000        & 20,000      & 20,000      \\
		Small FOV epochs                    & 557         & 115         & 115         \\
        Large FOV epochs                    & 162         & 120         & 118         \\
		Specificity                         & 98.8\%      & 99.2\%      & 99.8\%      \\
		Sensitivity (Synthetic Craters)     & 96.6\%      & 98\%        & 99.8\%      \\
		Sensitivity (True Craters)          & 6/27         & 6/27        & 18/27        
	\end{tabular}
    \caption{Comparison of different versions of our network. Specificity and synthetic crater sensitivity were determined from a testing set of 500 images that reflect the way in which the network was trained, as explained in the network training section, calculated with a threshold of 0.5.}
	\label{table:Versions}
\end{table}

\end{document}